\documentclass[prd,aps,showpacs,superscriptaddress,nofootinbib,preprint]{revtex4-1}
\usepackage{amsmath}
\usepackage{epsfig}
\usepackage{bm}  
\usepackage{ulem}
\usepackage[sans]{dsfont}  %
\usepackage{slashed}
\usepackage{color}

\allowdisplaybreaks

\begin{document}
\title{Mass Dependence of Higgs Production\\
 at Large Transverse Momentum\\  
 through a Bottom Quark Loop}

\author{Eric Braaten}
\email{braaten@mps.ohio-state.edu}
\affiliation{Department of Physics,
                  The Ohio State University,
                   Columbus, OH 43210, USA}
                                      
\author{Hong Zhang}
\email{zhang.5676@osu.edu}
\affiliation{Department of Physics,
                   The Ohio State University,  
                   Columbus, OH 43210, USA}

\author{Jia-Wei Zhang}
\email{jwzhang@cqust.edu.cn}
\affiliation{Department of Physics,
                   Chongqing University of Science and Technology,
                   Chongqing 401331, China}

%\pacs{12.38.Bx, 12.39.St, 13.85.Hd, 13.88.+e}

\date{\today}
                                  
\begin{abstract}
In the production of the Higgs through a bottom-quark loop, the transverse momentum distribution of the Higgs at large $P_T$ is complicated by its dependence on two other important scales: the bottom quark mass $m_b$ and the Higgs mass $m_H$. A strategy for simplifying the calculation of the cross section at large $P_T$ is to calculate only the leading terms in its expansion in $m_b^2/P_T^2$. In this paper, we consider the bottom-quark-loop contribution to the parton process $q\bar{q}\to H+g$ at leading order  in $\alpha_s$. We show that the leading power of $1/P_T^2$ can be expressed in the form of a factorization formula that separates the large scale $P_T$ from the scale of the masses. All the  dependence on $m_b$ and  $m_H$ can be factorized into a distribution amplitude for $b \bar b$ in the Higgs, a distribution amplitude for $b \bar b$ in a real gluon, and an endpoint contribution. The factorization formula can be used to organize the calculation of the leading terms in the expansion in $m_b^2/P_T^2$ so that every calculation involves at most two scales.
 
\end{abstract}

\maketitle

%%%%%%%%%%%%%%%%%%%%%%%%%%%%%%%%%%%%%%%%%%%
\section{Introduction}
\label{sec:intro}
%%%%%%%%%%%%%%%%%%%%%%%%%%%%%%%%%%%%%%%%%%%

The discovery of the Higgs boson in the year 2012 completed the Standard Model (SM) of particle physics \cite{Aad:2012tfa, Chatrchyan:2012ufa}. Many properties of the Higgs have since been measured, and they are in agreement with the theoretical predictions of the SM \cite{deFlorian:2016spz}. As the experimental precision improves with the collection of more and  more data at the Large Hadron Collider (LHC), it is important that theoretical uncertainties in the SM predictions are under control. The most straightforward way to reduce the theoretical uncertainties is to carry out calculations to higher orders in perturbation theory, and to resum to all orders logarithmic terms that spoil the perturbative expansion in certain kinematic regions. 

The dominant contribution to the cross section for Higgs production is through a top-quark loop due to the large Yukawa coupling $y_t=m_t/v$ of the top quark, where $m_t$ is the top-quark mass and $v=246$~GeV is the vacuum expectation value of the Higgs field. The contribution to the cross section from the interference between the bottom-quark loop and the top-quark loop can be as large as 8\% of the contribution from the top-quark loop \cite{Lindert:2017pky}. That contribution is suppressed by the Yukawa coupling $y_b=m_b/v$,  where $m_b$ is the bottom-quark mass. However the amplitude for the parton process $i j \to H+k$ from a bottom-quark loop at leading order (LO) in the QCD coupling constant $\alpha_s$ reveals the existence of double logarithms of  $m_H^2/m_b^2$ and $P_T^2/m_b^2$, where $m_H$ is the mass of the Higgs and $P_T$ is its transverse momentum. Since the High-Luminosity LHC will measure the Higgs $P_T$ distribution with a few percent accuracy, it is important to understand the bottom-quark-loop contribution to the same level of precision.  This requires calculating the process to higher orders in $\alpha_s$ and resumming large logarithms to all orders.

The $b$-quark-loop contribution to the Higgs $P_T$ distribution was first calculated in 1987 \cite{Ellis:1987xu,Baur:1989cm}, but a complete calculation at next-to-leading order (NLO) in $\alpha_s$ is still not available. The difficulty of the NLO calculation is mainly due to the existence of multiple scales, including the two mass scales $m_H$ and $m_b$ and the two kinematic scales $P_T$ and $\sqrt{\hat s}$, where $\hat s$ is the square of the parton center-of-mass energy. In the kinematic region where $m_b$ is much smaller than the other scales, the calculation can be simplified by taking the limit $m_b\to 0$. This limit is nontrivial because of  nonanalytic functions of $m_b$, such as $\log(P_T^2/m_b^2)$, that diverge in the $m_b\to 0$ limit. The $m_b\to 0$ limit can be calculated by solving the differential equations for master integrals in this limit \cite{Mueller:2015lrx}. This method has been used by Mueller and \"Ozt\"urk to calculate bottom-quark-loop contributions to the inclusive cross section for Higgs production to NLO \cite{Mueller:2015lrx}. The method has also been used by Melnikov, Tancredi, and Wever to calculate the helicity amplitudes for $i j \to H+k$ from a bottom-quark loop at NLO \cite{Melnikov:2016qoc,Melnikov:2017pgf}. Terms suppressed by powers of $m_b$ are neglected, and all logarithms of $m_b$ are included. As discussed in Ref.~\cite{Lindert:2017pky}, calculating the limit $m_b\to 0$ using differential equations for master integrals is very demanding of computing resources. A method for taking the limit at an earlier stage of the calculation would be desirable.

An NLO calculation does not necessarily produce a dramatic increase in accuracy. The relative error in an exclusive amplitude is probably order $\alpha_s^2 \log^4(P_T/m_b)$, and the relative error in a sufficiently inclusive cross section is order $\alpha_s^2 \log^2(P_T/m_b)$. To decrease the relative error to order $\alpha_s$ requires resumming the leading logarithms to all orders. The standard methods for resumming threshold logarithms  can be applied to large logarithms of $P_T/m_b$  with $P_T \ll m_b$. The sources of large logarithms of $m_b/P_T$ with $P_T \gg m_b$ are completely different. At leading order, the  large logarithms come from collinear regions of the loop momentum in which a $b$ and $\bar b$ have nearly collinear momenta and from soft regions in which a $b$ or $\bar b$ has momentum small compared to $P_T$. The effects of partial resummation of large logarithms of $m_b/P_T$ have been studied empirically without understanding their origin at higher orders \cite{Mantler:2012bj, Grazzini:2013mca, Banfi:2013eda, Bagnaschi:2015bop}. The resummation of logarithms in abelian QCD has been studied in Ref.~\cite{Melnikov:2016emg}. It would be useful to develop  systematic methods to  resum large logarithms from QCD radiative corrections.

Calculations to higher orders can be simplified and the resummation of logarithms can be facilitated by separating scales. An example is the Higgs Effective Field Theory (HEFT), in which the top quark mass $m_t$ is taken to be much larger than all other scales and the top quark is integrated out of the theory. Using HEFT, the  total cross section for Higgs production has been calculated to the impressive precision of N$^3$LO \cite{Anastasiou:2015ema, Anastasiou:2016cez}. The accuracy has been further improved by the resummation of threshold logarithms \cite{Ahrens:2008nc, Bonvini:2014joa, Li:2014afw, Bonvini:2014tea, Schmidt:2015cea, Bonvini:2016frm}. HEFT has also been used to calculate the cross section for Higgs plus one jet to $\text{N}^\text{2}\text{LO}$  \cite{Boughezal:2013uia, Chen:2014gva, Boughezal:2015dra, Boughezal:2015aha} and the cross section for Higgs plus two or more jets to NLO \cite{Campbell:2006xx, Campbell:2010cz, Cullen:2013saa}.

A  new approach to Higgs production at large transverse  momentum $P_T$ based on separation of scales has been introduced in Refs.~\cite{Braaten:2015ppa} and \cite{Braaten:2017lxx}. The separation of scales was accomplished by using factorization formulas that were deduced from factorization theorems for perturbative QCD. When there is a large kinematic scale $Q$, it is reasonable to expand in powers of $M^2/Q^2$, where $M$ represents scales provided by masses and nonperturbative low-energy scales. The expansion may not be straightforward because of terms that are nonanalytic in $M^2/Q^2$, such as logarithms of $M^2$ or functions of mass ratios. In Ref.~\cite{Braaten:2015ppa}, a factorization formula for the inclusive Higgs $P_T$ distribution at the leading power of $M^2/P_T^2$ was used to factor the nonanalytic terms into fragmentation functions. The factorization formula reproduces the LO result up to corrections of order $M^2/Q^2$, indicating  that theoretical  errors are under control at large $Q^2$. In the factorization approach, different energy scales are separated into different pieces in the factorization formula. Since fewer scales need to be considered in each piece, calculations to higher order are much simpler. The factorization formula also makes it possible to sum large logarithms of $M^2/P_T^2$ to all orders by solving evolution equations for the fragmentation functions.

In Ref.~\cite{Braaten:2017lxx}, we showed that the factorization approach can also be used to simplify amplitudes for exclusive Higgs production at large $P_T$. The specific example considered in  Ref.~\cite{Braaten:2017lxx} was the top-quark-loop contribution to the parton process $q\bar{q}\to H + g$ at LO. The relevant scales are the hard kinematic scales $Q\sim P_T, \sqrt{\hat s}$ and the soft mass scales $M\sim m_H, m_t$. The leading power in the expansion of the amplitude in powers of $M^2/Q^2$ was expressed in the form of a factorization formula in which the scales $M$ and $Q$ are separated. The factorization formula involves a distribution amplitude for a $t \bar t$ pair in the Higgs, a distribution amplitude for a $t \bar t$ pair in a real gluon, and an endpoint contribution from the transition $t +\bar t \to H + g$ via the exchange of a soft quark. The factorization formula provides a systematic approximation with errors of order $M^2/Q^2$ that go to zero as the kinematic scale $Q$ increases. Every piece in the factorization formula was calculated diagramatically in such a way that it involved either the scale $Q$ or the scale $M$. We also presented an improved factorization formula that includes all dependence on $m_t$ that is not suppressed by $m_H^2/Q^2$, so that the largest errors are reduced from order $m_t^2/Q^2$ to order $m_H^2/Q^2$.

In this paper, we apply the factorization method of Ref.~\cite{Braaten:2017lxx}  to the bottom-quark-loop contribution to the parton process $q\bar{q}\to H + g$  at LO. The relevant scales are the hard kinematic scales $Q\sim P_T, \sqrt{\hat s}$ and the soft mass scales $M\sim m_H, m_b$. The leading power in the expansion of the amplitude in powers of $M^2/Q^2$ is expressed in the form of a factorization formula in which the scales $M$ and $Q$ are separated. We also present an improved factorization formula that includes all dependence on $m_H$ that is not suppressed by $m_b^2/Q^2$, so that the largest errors are reduced from order $m_H^2/Q^2$ to order $m_b^2/Q^2$.

This paper is organized as follows. In Section~\ref{sec:FormFactor}, we introduce the form factor for the bottom-quark-loop contribution to the matrix element for $q\bar{q}\to H + g$. We define  the leading-power (LP) form factor to be the leading term in the expansion of the form factor in powers of $M^2/Q^2$. In Section~\ref{sec:RegularizedFF}, we calculate the LP form factor using dimensional regularization and rapidity regularization to regularize the divergences that arise from separating the contributions from different regions. In Section~\ref{sec:RenormalizedFF}, we renormalize all the ultraviolet divergences to obtain a finite factorization formula for the LP form factor. We also present an improved factorization formula with errors of order $m_b^2/Q^2$. In Section~\ref{sec:compare}, we show that the improved factorization formula gives a good approximation to the full form factor whose error decreases to 0 rapidly as $P_T$ increases. We discuss the prospects for extending our approach to NLO in $\alpha_s$ in Section~\ref{sec:Discuss}.

%%%%%%%%%%%%%%%%%%%%%%%%%%%%%%%%%%%%%%%%%%%
\section{Higgs production by $\bm{q \bar q \to H+g}$ through a $\bm{b}$ loop}
\label{sec:FormFactor}
%%%%%%%%%%%%%%%%%%%%%%%%%%%%%%%%%%%%%%%%%%%

In this  Section, we define the form factor that determines the bottom-quark-loop contribution to the cross section for $q \bar q\to H + g$ at leading order in $\alpha_s$. We give the leading power in the expansion of the form factor in powers of $M^2/Q^2$. We also present the schematic form of a factorization formula for the LP form factor.

%%%%%%%%%%%%%%%%%%%%%%%%%%%%%%%%
\subsection{Form factor for $\bm{g^* \to H+g}$} 
%%%%%%%%%%%%%%%%%%%%%%%%%%%%%%%%

%%%%%%%%%%%%%
\begin{figure}
\includegraphics[width=10cm]{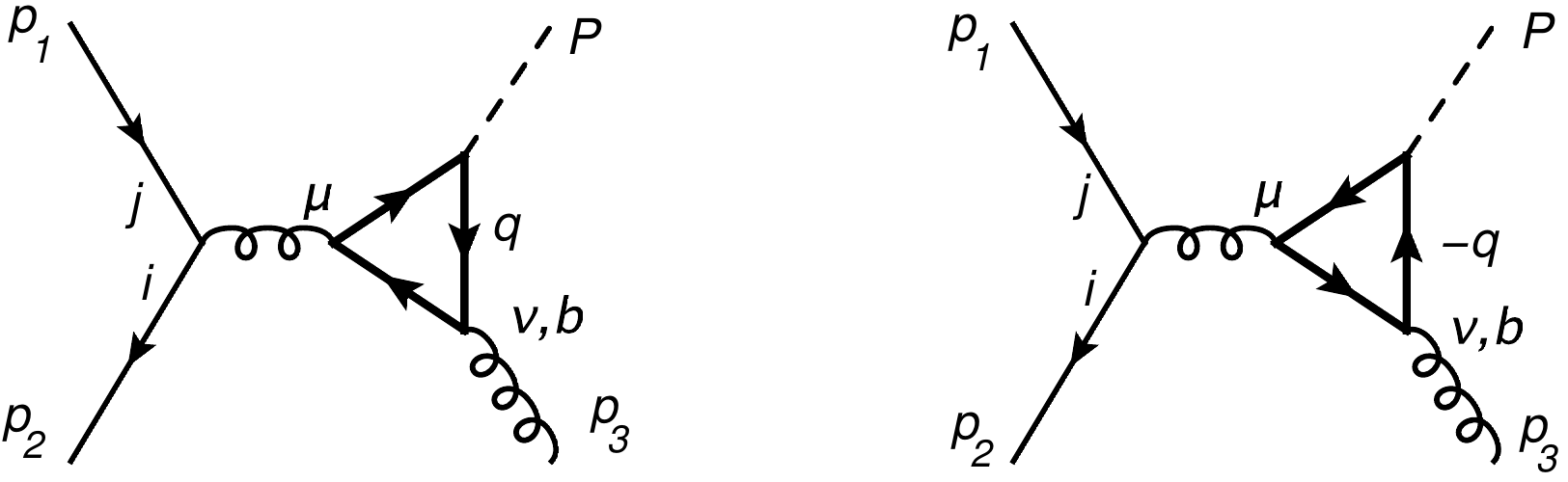}
\caption{Feynman diagrams for $q\bar{q}\to H+g$ through a bottom-quark loop at LO.
\label{fig:diagrams}}
\centering
\end{figure}
%%%%%%%%%%%%%

The reaction $q \bar q\to H + g$ proceeds at leading order (LO) in the QCD coupling constant $g_s$ through the two one-loop Feynman diagrams in Fig.~\ref{fig:diagrams}. The matrix element for $q(p_1) \bar q(p_2)\to H(P) + g(p_3)$ at LO has the form
%================
\begin{equation}
\mathcal{M}=  
\frac{g_s}{2\hat s} \, T^b_{ij}\,\bar v_2 \gamma_\mu u_1\, \mathcal{T}^{\mu \nu}(P,p_3)\, \varepsilon^*_{3\nu},
\label{eq:M}
\end{equation}
%================
where $T^b_{ij}$ is the color factor, $\bar v_2$ and $u_1$ are the Dirac spinors for $\bar q$ and $q$, and $\varepsilon_3$ is the polarization vector for the final-state gluon. The $q\bar{q}$ invariant mass $\hat{s}=(p_1+p_2)^2$ is also the invariant mass of the Higgs and the final-state gluon. The bottom-quark-loop contribution to the amplitude $\mathcal{T}^{\mu \nu}$ for $g^* \to H + g$ is
%================
\begin{equation}
\mathcal{T}^{\mu\nu} (P,p_3)= i g_s^2 y_b \int_q 
\frac{\text{Tr}\big[ ( \slash \!\!\!q + \slash \!\!\!\!P + m_b) \gamma^\mu 
(\slash \!\!\!q- \slash \!\!\!p_3 + m_b) \gamma^\nu (\slash \!\!\!q +m_b)\big]  - (m_b \to -m_b)}
{ [(q \!+\! P)^2 \!-\! m_b^2  \!+\! i \epsilon]\,  [q^2 \!-\! m_b^2  \!+\! i \epsilon]\,  
[(q \!-\! p_3)^2 \!-\! m_b^2  \!+\! i \epsilon]},
\label{eq:Tdef}
\end{equation}
%================
where the integration measure is $\int_q = \int d^4q/(2\pi)^4$. The color trace tr($T^a T^b$) has been absorbed into the prefactor of $\mathcal{T}^{\mu \nu}$ in Eq.~\eqref{eq:M}. The explicit Dirac trace in Eq.~\eqref{eq:Tdef} comes from the first diagram in Fig.~\ref{fig:diagrams}. Since the only nonzero terms in the trace are proportional to $m_b$ or $m_b^3$, the two diagrams are equal.

The Ward identities $(P+p_3)_\mu \mathcal{T}^{\mu \nu} = 0$ and $p_{3\nu} \mathcal{T}^{\mu \nu} = 0$ imply that the tensor $ \mathcal{T}^{\mu \nu}$ can be expressed in terms of two scalar form factors that are dimensionless functions of $\hat{s}$ and masses. Only one of the form factors contributes to the matrix element $\mathcal{M}$ in Eq.~\eqref{eq:M}. It can be expressed as
%================
\begin{equation}
\mathcal{F}(\hat s, m_b^2, m_H^2) = 
\frac{1}{(D-2)4 m_b}  \left( g_{\mu\nu} - \frac{p_{3\mu} (P+p_3)_\nu}{P.p_3} \right) \mathcal{T}^{\mu\nu} (P,p_3),
\label{eq:Fdef}
\end{equation}
%================
where $D=4$ is the number of space-time dimensions. The form factor can be expressed as an integral over a loop momentum:
%================
\begin{equation}
\mathcal{F}(\hat s, m_b^2, m_H^2) =  ig_s^2 y_b \int_q 
\frac{q^2 + 2 p_3.q + 2 P.p_3 + 3 m_b^2 - 4(P+p_3).q\, p_3.q/P.p_3}
{ [(q \!+\! P)^2 \!-\! m_b^2  \!+\! i \epsilon]\, [q^2 \!-\! m_b^2 \!+\! i \epsilon]\,
[(q \!-\! p_3)^2-m_b^2  \!+\! i \epsilon]}.
\label{eq:Fint4}
\end{equation}
%================
The square of the matrix element $\mathcal{M}$ for $q \bar q\to H + g$ 
summed over spins  and colors is proportional to $|\mathcal{F}|^2$: 
%================
\begin{equation}
\frac{1}{4N_c^2} \sum |\mathcal{M}|^2 =
\frac{2(N_c^2-1)g_s^2m_b^2}{N_c^2}\,
\frac{\hat{t}^2+\hat{u}^2}{\hat s(\hat{s}-m_H^2)^2}
\, |\mathcal{F}(\hat s, m_b^2, m_H^2)|^2.
\label{eq:MM-F}
\end{equation}
%================

The bottom-quark-loop contribution to the matrix elements for $g\,  q \to H + q$ and  $g \, \bar q \to H + \bar q$ at LO can be expressed in terms of the same function  $\mathcal{F}$  as the form factor for $q \bar q\to H + g$, but with the positive Mandelstam variable $\hat s$ replaced by a negative Mandelstam variable $\hat t$. If the form factor $\mathcal{F}$  for $q \bar q\to H + g$ is expressed in terms of the complex variable $\hat s + i \epsilon$, it can be applied to  $g\,  q \to H + q$ and $g\, \bar q \to H + \bar q$ by analytic continuation.

%%%%%%%%%%%%%%%%%%%%%%%%%%%%%%%%
\subsection{LP form factor}
\label{sec:LPFF}
%%%%%%%%%%%%%%%%%%%%%%%%%%%%%%%%

The form factor $\mathcal{F}$ is a function of the three energy scales $\hat s^{1/2}$, $m_b$, and $m_H$, which satisfy the inequalities $m_b < m_H \le \sqrt{\hat s}$. Analytic expressions for $\mathcal{F}$ are given in Refs.~\cite{Ellis:1987xu,Baur:1989cm}. The analytic expression for $\mathcal{F}$ can be simplified in the  limit  $m_b,m_H \ll\hat s^{1/2} $ by expanding in powers of $m_H^2/\hat s$ and  $m_b^2/\hat s$. We refer to the leading term in the expansion of the form factor in powers of $m_b^2/\hat s$ and $m_H^2/\hat s$ as the {\it leading-power (LP) form factor}. The LP form factor can be derived from the full  form factor in Refs.~\cite{Ellis:1987xu,Baur:1989cm}:
%================
\begin{eqnarray}
\mathcal{F}^\text{LP}(\hat s, m_b^2, m_H^2) &=& 
\frac{g_s^2 y_b}{32\pi^2} \left\{ - \log^2 \frac{-\hat s - i \epsilon}{m_b^2}  + 4\log \frac{-\hat s - i \epsilon}{m_b^2} \right.
+ \left( \log \frac{r + \sqrt{r^2-1}}{r - \sqrt{r^2-1}} - i \pi  \right)^2
\nonumber\\
&& \hspace{2cm}\left.
- \frac{4\sqrt{r^2-1}}{r} \left( \log \frac{r + \sqrt{r^2-1}}{r - \sqrt{r^2-1}} - i \pi  \right)
- 4 \right\},
\label{eq:FFLP-b}
\end{eqnarray}
%================
where $r$ is the mass ratio defined by
%================
\begin{equation}
r \equiv m_H/(2m_b).
\label{eq:r}
\end{equation}
%================
It can also be obtained from the top-quark loop contribution to the LP form factor in Ref.~\cite{Braaten:2017lxx} by analytically continuing the top quark mass $m_t$ to $m_b - i \epsilon$ and replacing the top-quark Yukawa coupling $y_t$ by $y_b$.

Another limit in which the analytic expression for $\mathcal{F}$ can be simplified is $m_b \ll m_H,\hat s^{1/2} $. The leading term in the expansion in powers of $m_b^2/m_H^2$ and  $m_b^2/\hat s$ depends logarithmically on $m_b$, and we must keep the $m_b$ dependence in the logarithms. The $m_b \to 0$ limit of the form factor is
%================
\begin{eqnarray}
\mathcal{F}(\hat s, m_b^2\to 0, m_H^2) &=& 
\frac{g_s^2 y_b}{32\pi^2} \left\{ 
-  \log^2\frac{-\hat{s}-i\epsilon}{m_H^2}
-  \left(2\log\frac{m_H^2}{m_b^2} -\frac{4\hat{s}}{\hat{s}-m_H^2}\right)\log\frac{\hat{s}+i\epsilon}{m_H^2}
- \pi^2 - 4 \right\}.
\nonumber\\
\label{eq:FFmb->0}
\end{eqnarray}
%================

%%%%%%%%%%%%%%%%%%%%%%%%%%%%%%%%
\subsection{Leading-power regions}
%%%%%%%%%%%%%%%%%%%%%%%%%%%%%%%%

The LP form factor in Eq.~\eqref{eq:FFLP-b} can be calculated directly using the {\it method of regions} \cite{Beneke:1997zp, Smirnov:2002pj}. There are four regions of the loop integral over the momentum $q$ in Eq.~\eqref{eq:Fint4} that contribute at leading power:
%%%%%%%%%%%%%
\begin{itemize}
\item
the {\it hard} region, in which $q^\mu$ is order $Q$, so $q^2$, $P.q$, and  $p_3.q$ are all order $Q^2$,
\item
the {\it Higgs collinear} region, in which $p_3.q$ is order $Q^2$, but $q^2$ and $P.q$ are order $M^2$,
\item
the {\it gluon collinear} region, in which $P.q$ is order $Q^2$ and $q^2$ and $p_3.q$ are order $M^2$,
\item
the {\it soft} region, in which $q^\mu$ is order $M$, so $q^2$ is order $M^2$ and $P.q$  and $p_3.q$ are order $MQ$.
\end{itemize}
%%%%%%%%%%%%%

The LP form factor $\mathcal{F}$ in  Eq.~\eqref{eq:FFLP-b} is finite, but the contributions from the individual leading-power regions have ultraviolet divergences and infrared divergences. The divergences cancel when all the  contributions are added. Some of the divergences can be regularized using dimensional regularization. The generalization of the integral in Eq.~\eqref{eq:Fint4} to $D=4-2\epsilon$ space-time dimensions can be obtained by the contraction  in Eq.~\eqref{eq:Fdef} of the tensor $\mathcal{T}^{\mu\nu}$ in Eq.~\eqref{eq:Tdef}. After evaluating the Dirac trace, the form factor reduces to
%================
\begin{eqnarray}
\mathcal{F}(\hat s, m_b^2, m_H^2) &=& \frac{2 ig_s^2 y_b}{D-2} \int_q 
\frac{1}
{ [(q \!+\! P)^2 \!-\! m_b^2  \!+\! i \epsilon]\,  [q^2 \!-\! m_b^2 \!+\! i \epsilon]\, [(q \!-\! p_3)^2-m_b^2  \!+\! i \epsilon]}
\nonumber\\
&&\hspace{-2cm}\times \left(
(5-D)q^2 - 4\frac{(P+p_3).q\, p_3.q}{P.p_3}
+  2(D-3)p_3.q + (D-2)P.p_3 + (D-1) m_b^2
\right),~
\label{eq:Fintdimreg}
\end{eqnarray}
%================
where the integration measure for the loop momentum is
%================
\begin{equation}
\int_q  \equiv \, \mu^{2\epsilon}
\frac{(4 \pi)^{-\epsilon}}{\Gamma(1+\epsilon)} \int \frac{d^D q}{(2 \pi)^D}.
\label{eq:intqD}
\end{equation}
%================
The regularized contributions to the LP form factor from each of the regions itemized above can be obtained from the expression for the integral in Eq.~\eqref{eq:Fintdimreg} by keeping only the leading terms at large $Q/M$ in the numerator and the leading terms in each of the denominators.

%%%%%%%%%%%%%%%%%%%%%%%%%%%%%%%%
\subsection{Factorization formula}
%%%%%%%%%%%%%%%%%%%%%%%%%%%%%%%%

In order to understand the dependence of the leading-power form factor in Eq.~\eqref{eq:FFLP-b} on the masses, it is necessary to separate the  dependence on $\hat s$ from the dependence on the masses $m_H$ and $m_b$. We refer to the kinematic scale $Q=\hat s^{1/2}$ as the {\it hard scale}. We refer to the scale $M$ provided by the masses $m_H$ and $m_b$ as the {\it soft scale}. The four regions itemized above correspond to four contributions to the LP form factor:
%%%%%%%%%%%%%
\begin{itemize}
\item
{\it direct production} of $H+g$, in which the Higgs $H$ and the real gluon $g$ are produced by the process $g^* \to H+g$ at the hard scale $Q$,
\item
$b \bar b$ {\it fragmentation} into $H$, in which a nearly collinear $b \bar b$ pair and the real gluon are created by the process $g^* \to b \bar b+g$ at the  hard scale $Q$, and the Higgs is produced by the subsequent transition $b \bar b \to H$ at the soft scale $M$,
\item
$b \bar b$ {\it fragmentation} into $g$, in which a nearly collinear $b \bar b$ pair and the Higgs are created by the process $g^* \to H +b \bar b$ at the  hard scale $Q$, and the real gluon is produced by the subsequent transition $b \bar b \to g$ at the soft scale $M$,
\item
{\it endpoint production} of $H+g$, in which a $b$ and $\bar b$ are created  by the process $g^* \to b + \bar b$ at the hard scale $Q$, and the Higgs and the real gluon are produced by the subsequent transition $b+ \bar b \to H + g$ at the soft scale $M$.
\end{itemize}
%%%%%%%%%%%%%

In Ref.~\cite{Braaten:2017lxx}, we showed that the top-quark-loop contribution to the LP form factor for $g^* \to H + g$ at LO can be expressed in terms of a factorization formula that separates the hard scale $Q$ from the masses $m_t$ and $m_H$. The analogous factorization formula for the bottom-quark-loop contribution to the LP form factor has the schematic form
%================
\begin{eqnarray}
\mathcal{F}^\text{LP}[H+g] &=&  \widetilde{\mathcal{F}} [H+g]   
+ \widetilde{\mathcal{F}}[b \bar b_{1V}+g]  \otimes d[b \bar b_{1V} \to H]
\nonumber\\
&& \hspace{2cm}
+ \widetilde{\mathcal{F}} [H + b \bar b_{8T}]  \otimes d[b \bar b_{8T} \to g]
+ \mathcal{F}_\text{endpt}[H+g].
\label{eq:Ffact}
\end{eqnarray}
%================
The terms on the right side correspond to the  four contributions itemized above. The subscripts on $b \bar b$ indicate the color channel, which can be color-singlet (1) or color-octet (8), and the Lorentz channel, which can be vector ($V$) or tensor ($T$). The $\otimes$ represents an integral over the relative longitudinal momentum fraction $\zeta$ of the $b \bar b$ pair, whose range is $-1 \le \zeta \le +1$. The factors represented by $\widetilde{\mathcal{F}}$ are {\it hard form factors} that depend only on the hard scale $Q$. The factors represented by $d$ are {\it distribution amplitudes} that depend only on the soft scale $M$. Regularized expressions for each of the pieces in the factorization formula in Eq.~\eqref{eq:Ffact} will be obtained in Section~\ref{sec:RegularizedFF}. Renormalized expressions for each of the pieces in the factorization formula will be given in Section~\ref{sec:RenormalizedFF}.

%%%%%%%%%%%%%%%%%%%%%%%%%%%%%%%%%%%%%%%%%%%
\section{Regularized Factorization Formula}
\label{sec:RegularizedFF}
%%%%%%%%%%%%%%%%%%%%%%%%%%%%%%%%%%%%%%%%%%%

In this Section, we calculate each of the pieces in the factorization formula for the LP form factor in Eq.~\eqref{eq:Ffact} in a way that involves only the single scale $Q$ or $M$. We use dimensional regularization and rapidity regularization to regularize the divergences in the contributions to the LP form factor from each of the leading-power regions. The divergences cancel when the four terms on the right side of Eq.~\eqref{eq:Ffact} are added.

%%%%%%%%%%%%%%%%%%%%%%%%%%%%%%%%
\subsection{Rapidity regularization and zero-bin subtraction}
%%%%%%%%%%%%%%%%%%%%%%%%%%%%%%%%

The contributions to the LP form factor from the individual leading-power regions have ultraviolet divergences and infrared divergences. Some of the divergences are regularized by the dimensional regularization of the loop integral in Eq.~\eqref{eq:Fintdimreg}. There are additional infrared divergences called {\it rapidity divergences} that require some other regularization procedure. We regularize the rapidity divergences using a method called {\it rapidity regularization}. Rapidity regularization in conjunction with zero-bin subtraction  was introduced as a method for regularizing rapidity divergences by Manohar and Stewart \cite{Manohar:2006nz}. Rapidity regularization separates the contributions from collinear and soft regions by explicitly breaking the boost invariance. Zero-bin subtractions of collinear contributions are required to avoid double counting of soft contributions. With rapidity regularization and zero-bin subtraction, the rapidity divergence from each region is an ultraviolet divergence. This allows the cancellation of rapidity divergences to be implemented as a renormalization procedure.

In order to specify the rapidity regularization factors, it is convenient to introduce light-like vectors $n$ and $\bar n$ such that the only components of $P^\mu$ and $p_3^\mu$ that are of order $Q$ are $P.n$ and $p_3.\bar n$. We choose the normalizations of $n$ and $\bar n$  so that $n . \bar n=2$, which implies $P.n\, p_3.\bar n = \hat s$. Dimensional regularization is used to separate the hard contribution from the sum of the remaining contributions. The integration measure of the loop momentum in Eq.~\eqref{eq:intqD} can be expressed as
%================
\begin{equation}
\int_q \equiv \int \frac{d (q.n) d(q.\bar n)}{8 \pi^2} \int_{\bm{q}_\perp},
\label{eq:intq}
\end{equation}
%================
where the measure of the dimensionally regularized transverse momentum integral is
%================
\begin{equation}
\int_{\bm{q}_\perp}  \equiv \, \mu^{2\epsilon}
\frac{(4 \pi)^{-\epsilon}}{\Gamma(1+\epsilon)} \int \frac{d^{2-2\epsilon}q_\perp}{(2 \pi)^{2-2\epsilon}}.
\label{eq:intqperp}
\end{equation}
%================
We can use the 4-vectors $n$ and $\bar n$ to define regions of the loop momentum $q$. In the $n$ collinear region, $q.\bar n$ is order $Q$, $q^2$ is order $M^2$, and $q.n$ is order $M^2/Q$. In the $\bar n$ collinear region, $q.n$ is order $Q$, $q^2$ is order $M^2$, and $q.\bar n$ is order $M^2/Q$. In the soft region, $q. n$, $q.\bar n$, and $\bm{q}_\perp$ are all order $M$.

With rapidity regularization, different regularization factors may be used in different regions. The specific forms of the regularization factors required for our problem were used in Ref.~\cite{Chiu:2011qc} and described more explicitly in Ref.~\cite{Chiu:2012ir}. The regularization factors in each of the regions of $q$ are
%================
\begin{subequations}
\begin{eqnarray}
n~\text{collinear:} ~&&~  \big( |q.\bar n|/\nu_- \big)^{-\eta},
\label{eq:rapidreg-n}
\\\
\bar n~\text{collinear:} ~&&~ \big( |q.n|/\nu_+ \big)^{-\eta},
\label{eq:rapidreg-nbar}
\\\
\text{soft:}~~~~~~~~~ ~&&~ \big( |q.(n-\bar n)|/\nu \big)^{-\eta},
\label{eq:rapidreg-soft}
\end{eqnarray}
\label{eq:rapidreg}%
\end{subequations}
%================
where $\eta$ is the regularization parameter and $\nu_+$, $\nu_-$, and $\nu$ are regularization scales. The term $ |q.(n-\bar n)|$ in the soft factor reduces to $ |q.\bar n|$ in the $n$ collinear region and to $|q. n|$ in the $\bar n$ collinear region, so the essential difference between the three factors in Eq.~\eqref{eq:rapidreg} is in the regularization scales. They are constrained by an equation that depends on the application. In most previous cases, the equation  was either $\nu_+ \nu _- = \nu^2$ or $\nu_+ \nu _- = -\nu^2$.

%%%%%%%%%%%%%%%%%%%%%%%%%%%%%%%%
\subsection{Hard contribution}
\label{sec:Fhard0}
%%%%%%%%%%%%%%%%%%%%%%%%%%%%%%%%

The contribution to the  LP form factor from the hard region in which $q^\mu$ is order $Q$ is 
%================
\begin{eqnarray}
\mathcal{F}^{\text{LP}}_\text{hard}(\hat s) &=& 
 \frac{2 ig_s^2 y_b}{D-2} \int_q \frac{1}{[(q + \tilde P)^2  + i \epsilon]\,  [q^2 + i \epsilon]\, [(q - p_3)^2 + i \epsilon]}
\nonumber\\
&&\hspace{1cm}\times 
\left( (5-D)q^2 - 4\frac{(\tilde P+p_3).q\, p_3.q}{\tilde P.p_3}+  2(D-3)p_3.q + (D-2)\tilde P.p_3 \right).
\label{eq:F0hardint}
\end{eqnarray}
%================
The 4-momentum $P$ of the Higgs has been replaced by a light-like 4-vector $\tilde P$ whose 3-vector component is collinear to $\bm{P}$ and whose normalization is given by $2\tilde{P}.p_3=\hat{s}$. The integral in Eq.~\eqref{eq:F0hardint} can be calculated analytically. A Laurent expansion in $\epsilon$ gives
%================
\begin{equation}
\mathcal{F}^{\text{LP}}_\text{hard}(\hat s) = 
- \frac{g_s^2 y_b}{16 \pi^2} \left[\frac{-\hat s - i \epsilon}{\mu^2}\right]^{-\epsilon}
\left(  \frac{1}{\epsilon^2} +\frac{2}{\epsilon} + 6 - \frac{\pi^2}{6}   \right).
\label{eq:F0hardeps}
\end{equation}
%================
The poles in $\epsilon$ have an infrared origin, but they can be transformed into ultraviolet poles in $\epsilon$ by adding integrals that have no scale and therefore vanish with dimensional regularization.

%%%%%%%%%%%%%%%%%%%%%%%%%%%%%%%%
\subsection{Higgs collinear contribution}
%%%%%%%%%%%%%%%%%%%%%%%%%%%%%%%%

As shown in Ref.~\cite{Braaten:2017lxx}, the scales $Q$ and $M$ in the Higgs collinear contribution to the LP form factor can be separated by expressing it  as an integral over the relative longitudinal momentum fraction $\zeta$ of the collinear $b$ and $\bar b$ that form the Higgs:
%================
\begin{equation}
\mathcal{F}_{H\, \text{coll}}^{\text{LP}} =
 \int_{-1}^{+1}\!\!\!\! d \zeta \,\mathcal{\widetilde F}_{b \bar b_{1V}+ g}(\zeta) \,
d_{b \bar b_{1V} \to H}(\zeta) .
\label{eq:FLPHcoll1V}
\end{equation}
%================
The integrand is the product of a hard form factor $\mathcal{\widetilde F}_{b \bar b_{1V}+ g}$ for producing a collinear $b \bar b$ pair in the color-singlet Lorentz-vector ($1V)$ channel plus a gluon and a distribution amplitude $d_{b \bar b_{1V} \to H}$ for a $b \bar b$ pair in the Higgs. The hard form factor depends only on the scale $Q$. The distribution amplitude depends on the scale $M$. With rapidity regularization, it also depends logarithmically on $P.n$. 

%%%%%%%%%%%%%
\begin{figure}
\includegraphics[width=10cm]{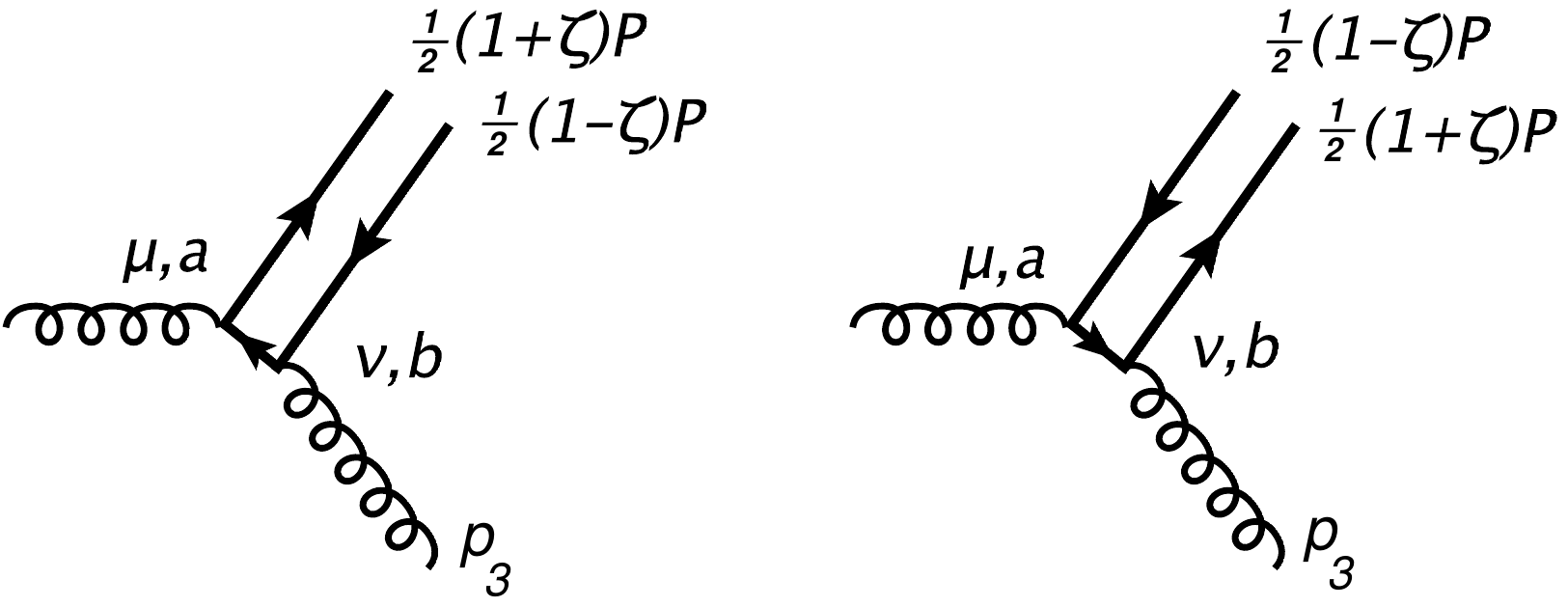}
\caption{Feynman diagrams for  the tensor amplitude $\mathcal{T}^{\mu\nu}$ for $g^*\to b\bar b +g$ at LO.
\label{fig:tt+g}}
\centering
\end{figure}
%%%%%%%%%%%%%

The hard form  factor in Eq.~\eqref{eq:FLPHcoll1V} is derived from the amplitude $\mathcal{T}^{\mu  \nu}$ for $g^* \to b \bar b + g$, which is given by the sum of the two diagrams in Fig.~\ref{fig:tt+g}. Since we only want the leading power, we can set $m_b=0$. We need the amplitude for producing $b$ and $ \bar b$  in the color-singlet Lorentz-vector ($1V$) channel and with collinear momenta $\tfrac12(1+\zeta) \tilde P$ and $\tfrac12(1-\zeta) \tilde P$ plus a real gluon with momentum $p_3$. The $b \bar b$ pair can be projected onto the color-singlet state by tracing over the color indices of $b$ and $\bar b$ and dividing by $\sqrt{N_c}$, where $N_c=3$ is the number of quark colors. The $b \bar b$ pair can be projected onto the Lorentz-vector channel by replacing the outer product $v \, \bar u$ of the $\bar b$ and $b$ spinors by $ \slash \!\!\!\! \tilde P$. The $1V$ contribution to the tensor amplitude is  
%================
\begin{equation}
\mathcal{T}_{1V}^{\mu\nu} (P,p_3)= -\frac{4g_s^2}{\sqrt{N_c}} 
\left( \frac{\tilde P.p_3  g^{\mu \nu} - (\tilde P^\mu p_3^\nu +  p_3^\mu \tilde P^\nu ) 
- (1-\zeta) \tilde P^\mu  \tilde P^\nu }
{(1-\zeta)\tilde P.p_3 } - (\zeta \to -\zeta) \right).
\label{eq:T-bb1V+g}
\end{equation}
%================

The hard form factor for $g^* \to b \bar b_{1V} + g$ is obtained by the contraction in Eq.~\eqref{eq:Fdef} of the tensor $\mathcal{T}_{1V}^{\mu\nu}$ in Eq.~\eqref{eq:T-bb1V+g}, with $P$ replaced by $\tilde P$. We choose to move a factor $1/(1-\zeta^2)$ from the hard form factor to the distribution amplitude to allow the poles in the regularization parameters to be made explicit. A canceling factor $1-\zeta^2$ must appear in the hard form factor. We choose to also move the factor  $1/(\sqrt{N_c} \, m_b)$ to the distribution amplitude to simplify the expressions for both the hard form factor and the distribution amplitude. The resulting expression for the hard form factor is
%================
\begin{equation}
\widetilde{\mathcal{F}}_{b \bar b_{1V}+ g}(\zeta) = - 2 g_s^2 \zeta.
\label{eq:FFbb1V+g}
\end{equation}
%================

The distribution amplitude for $b \bar b_{1V} \to H$ in Eq.~\eqref{eq:FLPHcoll1V} is a function of the relative longitudinal momentum fraction $\zeta$ that describes how the longitudinal momentum of the Higgs is  distributed between a $b$ and a $\bar b$. The distribution amplitude can be calculated by using ingredients from the Feynman rules for double-parton fragmentation functions in Ref.~\cite{Ma:2013yla}. A fragmentation function can be expressed as the sum of cut diagrams that are products of an amplitude and the complex conjugate of an amplitude. The amplitude for $b \bar b$ fragmentation into a specific final state is the amplitude for that final state to be produced by sources that create the $b$ and the $\bar b$ in a specified color and Lorentz channel with relative longitudinal momentum fraction $\zeta$. The sources are the endpoints of eikonal lines that extend to future infinity. The Feynman rule for the sources that create the $b \bar b$ pair in the $1V$ channel is the product of a color matrix, a Dirac matrix, and a delta function
that are given in Ref.~\cite{Braaten:2017lxx}.

%%%%%%%%%%%%%
\begin{figure}
\includegraphics[width=4cm]{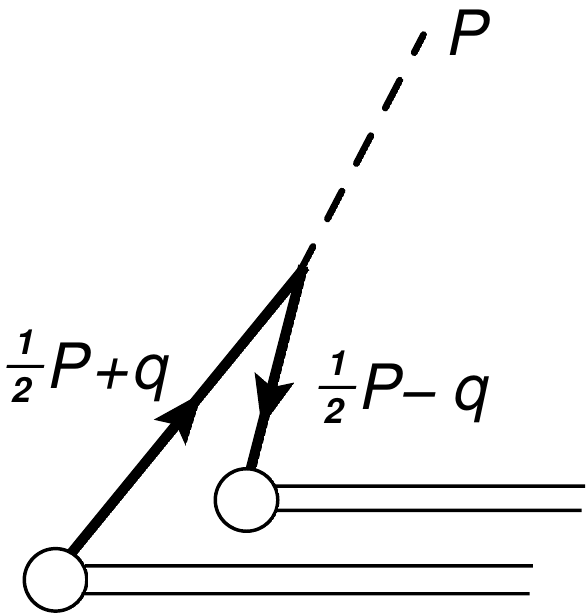}
\caption{Feynman diagram for the distribution amplitude for $b \bar b_{1V} \to H$ at LO.
\label{fig:tt->H}}
\centering
\end{figure}
%%%%%%%%%%%%%

The leading-order diagram for the distribution amplitude for a $b \bar b$ pair  in the Higgs is shown in Fig.~\ref{fig:tt->H}. The expression for the distribution amplitude is $\sqrt{N_c} \,y_bm_b\, \zeta \, d(\zeta)$, where the function $d(\zeta)$ is
%================
\begin{equation}
d(\zeta) = -i  \int_q  
\frac{ \delta(\zeta - 2q.n/P.n) }
{[(\tfrac12 P+q)^2 - m_b^2  + i \epsilon]\, [(\tfrac12 P-q)^2-m_b^2  + i \epsilon] }.
\label{eq:drapid}
\end{equation}
%================
We have suppressed the rapidity regularization factor and zero-bin subtractions for the integral over the loop momentum $q$. The rapidity regularization factor is the product of two factors like that in Eq.~\eqref{eq:rapidreg-nbar} with $q$ replaced by $\tfrac12P+q$ and by $\tfrac12P-q$. Multiplying by the factors $1/(\sqrt{N_c} \,m_b)$ and $1/(1-\zeta^2)$ that were removed from the form factor for $g^* \to b \bar b_{1V} + g$ in Eq.~\eqref{eq:FFbb1V+g}, we obtain the distribution amplitude 
%================
\begin{equation}
d_{b \bar b_{1V} \to H}(\zeta) = y_b \, \zeta \, \frac{d(\zeta)}{1-\zeta^2} .
 \label{eq:dfragbb1V->H}
\end{equation}
%================
The function $d(\zeta)$ is calculated with rapidity regularization and with appropriate zero-bin subtractions in the appendix of  Ref.~\cite{Braaten:2017lxx}. It has ultraviolet divergences that can be made explicit as poles in the regularization parameters $\epsilon$ and $\eta$ if the function is divided by $1-\zeta^2$:
%================
\begin{eqnarray}
 \frac{d(\zeta)}{1-\zeta^2}  =
\frac{1}{32 \pi^2}
\left[\frac{\mu^2}{m_b^2}\right]^\epsilon \left[\frac{P.n}{\nu_1}\right]^{-2\eta}
\frac{1}{\epsilon}
\left( -\frac{1}{2\eta}\delta(1-\zeta^2)+\frac{1}{(1-\zeta^2)_+}\right)
\nonumber\\
\times
\left[1-(1-\zeta^2)r^2-i\epsilon\right]^{-\epsilon}.
\label{eq:d-analreg}
\end{eqnarray}
%================
We have set the rapidity regularization scale to $\nu_1$. The plus distribution can be defined by giving the integral of the product of the distribution and a smooth function $f(z)$ over the closed interval $-1 \le z \le +1$:
%================
\begin{equation}
\int_{-1}^{+1}\!\!d \zeta \, g(\zeta)_+ \, f(\zeta) \equiv
\int_{-1}^{+1}\!\!d \zeta \, g(\zeta)  \,  \frac{f(\zeta)+f(-\zeta)-f(1)-f(-1)}{2}.
 \label{eq:intplus}
\end{equation}
%================
The distribution amplitude in Eq.~\eqref{eq:dfragbb1V->H} is 
%================
\begin{eqnarray}
d_{b \bar b_{1V} \to H}(\zeta) =
\frac{y_b}{32 \pi^2}
\left[\frac{\mu^2}{m_b^2}\right]^\epsilon \left[\frac{P.n}{\nu_1}\right]^{-2\eta}
\zeta  \Bigg\{\frac{1}{\epsilon}
\left( -\frac{1}{2\eta}\delta(1-\zeta^2)+\frac{1}{(1-\zeta^2)_+}\right) 
\nonumber\\
- \frac{\log\big(1- (1 - \zeta^2) r^2-i\epsilon \big)}{1 - \zeta^2} \Bigg\}.
\label{eq:dfragbb1V->Hreg}
\end{eqnarray}
%================

The Higgs collinear contribution to the LP form factor is obtained by evaluating the integral over $\zeta$ in Eq.~\eqref{eq:FLPHcoll1V}:
%================
\begin{eqnarray}
\mathcal{F}_{H\, \text{coll}}^\text{LP}(m_b^2,\tilde P.n)  &= &
\frac{g_s^2 y_b}{16 \pi^2} \left[ \frac{\mu^2}{m_b^2} \right]^{\epsilon}  
\Bigg\{\frac{1}{\epsilon}
\left(\frac{1}{2\eta}-\log\frac{\tilde P.n}{\nu_1}+2 \right) 
\nonumber \\
&&\hspace{3cm} 
+  \int_{-1}^{+1}\!\!\! d\zeta \, \zeta^2 \frac{\log\big(1- (1 - \zeta^2) r^2-i\epsilon \big)}{1 - \zeta^2} \Bigg\}.
\label{eq:FHcollint1}
\end{eqnarray}
%================
It  depends logarithmically on $\tilde P.n$. The remaining integral over $\zeta$ is
%================
\begin{eqnarray}
\int_{-1}^{+1}\!\!\! d\zeta \, \zeta^2 \frac{\log\big(1- (1 - \zeta^2) r^2-i\epsilon \big)}{1 - \zeta^2}   &=&
\frac12 \left( \log \frac{r + \sqrt{r^2-1}}{r - \sqrt{r^2-1}} - i \pi  \right)^2
\nonumber\\
&&\hspace{0cm}
- \frac{2\sqrt{r^2-1}}{r} \left( \log \frac{r + \sqrt{r^2-1}}{r - \sqrt{r^2-1}} - i \pi  \right)+4.
\label{eq:intzeta}
\end{eqnarray}
%================

%%%%%%%%%%%%%%%%%%%%%%%%%%%%%%%%
\subsection{Gluon collinear contribution}
%%%%%%%%%%%%%%%%%%%%%%%%%%%%%%%%

As shown in Ref.~\cite{Braaten:2017lxx}, the scales $Q$ and $M$ in the gluon collinear contribution to the LP form factor  can be separated by expressing it  as an integral over the relative longitudinal momentum fraction $\zeta$:
%================
\begin{equation}
\mathcal{F}_{g\, \text{coll}}^{\text{LP}} =
 \int_{-1}^{+1}\!\!\!\! d \zeta \,\mathcal{\widetilde F}_{H+ b \bar b_{8T}}(\zeta) \,
d_{ b \bar b_{8T} \to g}(\zeta) .
\label{eq:FLPgcoll8T}
\end{equation}
%================
The integrand is the product of the hard form factor $\mathcal{\widetilde F}_{H+ b \bar b_{8T}}$ for producing a Higgs plus a collinear $b \bar b$ pair in the color-octet Lorentz-tensor ($8T)$ channel and the  distribution amplitude $d_{b \bar b_{8T} \to g}$ for a $b \bar b$ pair in a real gluon. The hard form factor depends only on the scale $Q$. The distribution amplitude depends on the scale $M$. With rapidity regularization, it also depends logarithmically on $p_3.\bar n$. 

%%%%%%%%%%%%%
\begin{figure}
\includegraphics[width=9cm]{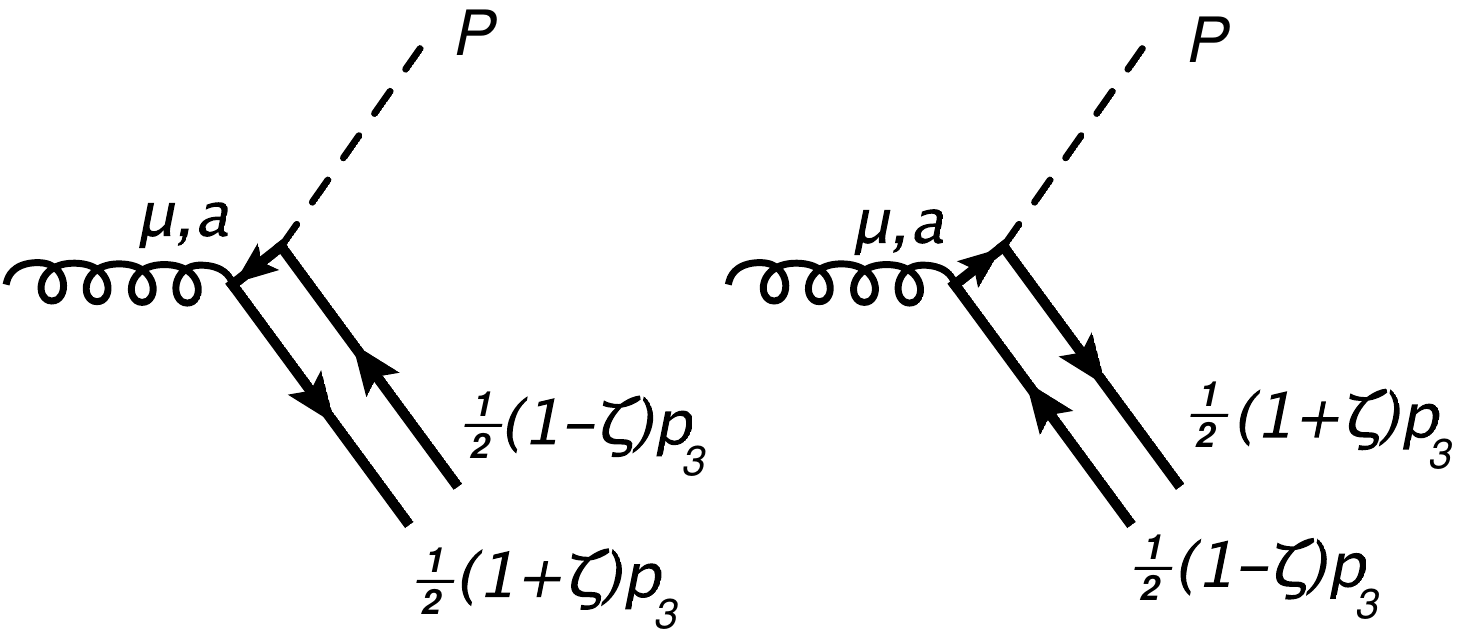}
\caption{Feynman diagrams for the tensor amplitude $\mathcal{T}^{\mu a,b}$ for $g^*\to H+ b\bar b$ at LO.
\label{fig:H+tt}}
\centering
\end{figure}
%%%%%%%%%%%%%

The hard form factor in Eq.~\eqref{eq:FLPgcoll8T} is determined from the amplitude $\mathcal{T}^\mu$ for $g^* \to H+  b \bar b$, which is given by the sum of the two diagrams in Fig.~\ref{fig:H+tt}. Since we only want the leading power, we can set $m_b = 0$. We need the amplitude for producing $b$ and $\bar b$ in the color-octet Lorentz-tensor ($8T$) channel with collinear momenta $\tfrac12(1+\zeta)  p_3$ and $\tfrac12(1-\zeta) p_3$ plus a Higgs with momentum $\tilde P$. The $b \bar b$ pair can be projected onto a color-octet state with color index $a$ by tracing the amplitude with the color matrix $\sqrt{2}\,T^a$. The $b \bar b$ pair can be projected onto the Lorentz-tensor channel with a Lorentz index $\nu$ by replacing the outer spinor product $v \, \bar u$ by $\slash \!\!\!\! p_3 \gamma_\perp^\nu$, where $ \gamma_\perp^\nu = g_\perp^{\nu\alpha} \gamma_\alpha$ are Dirac matrices that are perpendicular to $n$ and $\bar n$ and $g_\perp^{\nu\alpha}$ is the perpendicular  metric tensor:
%================
\begin{equation}
 g_{\perp\alpha\beta} = 
 g_{\alpha\beta} - \frac{n_\alpha \bar n_\beta + \bar n_\alpha n_\beta}{n.\bar n}.
\label{eq:gperp}
\end{equation}
%================
The $8T$ contribution to the vector amplitude $\mathcal{T}^{\mu}$ defines the tensor amplitude
%================
\begin{equation}
\mathcal{T}_{8T}^{\mu\nu} (P,p_3)= - 4 \sqrt{2}g_s y_b\,  g_\perp^{\mu\nu}
 \left(\frac{1}{1+\zeta} + \frac{1}{1-\zeta}  \right).
\label{eq:T-H+tt8T}
\end{equation}
%================

The hard form factor for $g^* \to H+ b \bar b_{8T}$  can be obtained  by the contraction in Eq.~\eqref{eq:Fdef} of the tensor $\mathcal{T}_{8T}^{\mu\nu}$ in Eq.~\eqref{eq:T-H+tt8T}, with $P$ replaced by $\tilde P$. We choose to move a factor $1/(1-\zeta^2)$ to the distribution amplitude to allow the poles in the regularization parameters to be made explicit. A canceling factor $1-\zeta^2$ must appear in the hard form factor. We also choose to move the factor $\sqrt{2}/m_b$ to the distribution amplitude to simplify the expressions for both the hard form factor and the distribution amplitude. The resulting expression for the hard form factor is  
%================
\begin{equation}
\widetilde{\mathcal{F}}_{H+ b \bar b_{8T}}(\zeta) = - 2 g_s y_b.
\label{eq:FFH+bb8T}
\end{equation}
%================

\begin{figure}
\includegraphics[width=4cm]{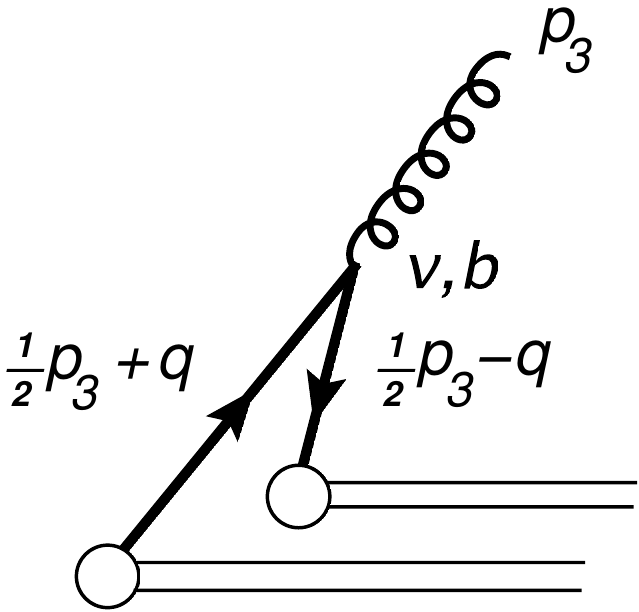}
\caption{Feynman diagram for the distribution amplitude for ${b \bar b_{8T} \to g}$ at LO.
\label{fig:tt->g}}
\centering
\end{figure}

The distribution amplitude for $b \bar b_{8T} \to g$ in Eq.~\eqref{eq:FLPgcoll8T} is  a function of the relative  longitudinal momentum fraction $\zeta$ that describes how the longitudinal momentum of the real gluon is  distributed between a $b$ and a $\bar b$. It can be calculated from the diagram  in Fig.~\ref{fig:tt->g} by using ingredients from the  Feynman rules for double-parton fragmentation functions in Ref.~\cite{Ma:2013yla}. The Feynman rule for the sources that create the $b \bar b$ pair in the $8T$ channel is the product of a color matrix, a Dirac matrix, and a delta function that are given in Ref.~\cite{Braaten:2017lxx}.

%%%%%%%%%%%%%
%%%%%%%%%%%%%

The leading-order diagram for the distribution amplitude for a $b \bar b$ pair in a real gluon is shown  in Fig.~\ref{fig:tt->g}. The amplitude for the source to  produce a real gluon with transverse polarization vector in the same direction as the source  and with the same color index as the source is $(g_s m_b/\sqrt{2}) d_0(\zeta)$, where the function $d_0(\zeta)$ is
%================
\begin{eqnarray}
d_0(\zeta) = -i \int_q  
\frac{ \delta(\zeta - 2q.\bar n/p_3.\bar n)}
{[(\tfrac12 p_3+q)^2 - m_b^2  + i \epsilon]\, [(\tfrac12 p_3-q)^2-m_b^2  + i \epsilon] }.
\label{eq:d0rapid}
\end{eqnarray}
%================
We have suppressed rapidity regularization factors and zero-bin subtractions in the integral over the loop momentum $q$. Multiplying by the factors $\sqrt{2}/m_b$ and $1/(1-\zeta^2)$ that were removed from the form factor for $g^* \to H+b \bar b_{8T} $, we obtain the distribution amplitude 
%================
\begin{equation}
d_{b \bar b_{8T} \to g}(\zeta) = g_s \, \frac{d_0(\zeta)}{1-\zeta^2} .
 \label{eq:dfragtt8T->g}
\end{equation}
%================
The function $d_0(\zeta)/(1-\zeta^2)$  with rapidity regularization can be obtained from the function $d(\zeta)/(1-\zeta^2)$ in Eq.~\eqref{eq:d-analreg}  by setting  $r=0$ and replacing $P.n$ with $p_3.\bar n$. The distribution amplitude in Eq.~\eqref{eq:dfragtt8T->g} is
%================
\begin{eqnarray}
d_{b \bar b_{8T} \to g}(\zeta) =
\frac{g_s}{32 \pi^2}
\left[\frac{\mu^2}{m_b^2}\right]^\epsilon \left[\frac{p_3.\bar n}{\nu_3}\right]^{-2\eta}
\frac{1}{\epsilon} \left\{-\frac{1}{2\eta}\delta(1-\zeta^2)+\frac{1}{(1-\zeta^2)_+}\right\}.
\label{eq:dfragtt8T->greg}
\end{eqnarray}
%================
We have set the rapidity regularization scale to $\nu_3$. The poles in $\epsilon$ and $\eta$ have ultraviolet origins. The plus distribution is defined in Eq.~\eqref{eq:intplus}.

The gluon collinear contribution to the LP form factor is obtained by evaluating the integral over $\zeta$ in Eq.~\eqref{eq:FLPgcoll8T}:
%================
\begin{equation}
\label{eq:Fgcollrap}
\mathcal{F}^{\text{LP}}_{g\,\text{coll}}(m_b^2,p_3.\bar n)
= \frac{g_s^2 y_b}{16\pi^2}
\left[\frac{\mu^2}{m_b^2}\right]^\epsilon
\frac{1}{\epsilon}
\left(\frac{1}{2\eta}-\log\frac{p_3.\bar n}{\nu_3} \right).
\end{equation}
%================
It depends logarithmically on $p_3.\bar n$.

%%%%%%%%%%%%%%%%%%%%%%%%%%%%%%%%
\subsection{Soft contribution}
%%%%%%%%%%%%%%%%%%%%%%%%%%%%%%%%

In the soft region of the loop momentum $q$, all the components of $q$ are order $M$. The soft contribution to the LP form factor  with rapidity regularization is
%================
\begin{equation}
\label{eq:Fsoftrapint}
\mathcal{F}^{\text{LP}}_{\text{soft}}
=2i g_s^2 y_b 
\int_q\frac{\tilde P.p_3}{[2\tilde P.q +i\epsilon]\, [q^2-m_b^2+i\epsilon]\, [-2p_3.q+i\epsilon]}
 \left[\frac{|q.(n-\bar n)|}{\nu} \right]^{-2\eta}.
\end{equation}
%================
The 4-momentum $P$ of the Higgs has been replaced by the light-like 4-vector $\tilde P$. The rapidity divergences from the two denominators $2\tilde P.q$ and $-2p_3.q$ have been regularized by multiplying the integrand by two identical copies of the factor in Eq.~\eqref{eq:rapidreg-soft}. The integral over $q$ in Eq.~\eqref{eq:Fsoftrapint} gives ultraviolet  poles in $\eta$ and in $\epsilon + \eta$. After a Laurent expansion in $\eta$,  the   soft contribution  reduces to
%================
\begin{equation}
\label{eq:Fsoftrap}
\mathcal{F}^{\text{LP}}_{\text{soft}}(m_b^2) =
- \frac{g_s^2 y_b}{16\pi^2}
\left[\frac{\mu^2}{m_b^2}\right]^\epsilon
\left(\frac{1}{\epsilon\, \eta} -\frac{1}{\epsilon^2}+\frac{1}{\epsilon}\log\frac{\nu^2}{m_b^2} +\frac{\pi^2}{6} \right).
\end{equation}
%================

%%%%%%%%%%%%%%%%%%%%%%%%%%%%%%%%
\subsection{LP form factor}
%%%%%%%%%%%%%%%%%%%%%%%%%%%%%%%%

In the sum of the Higgs collinear contribution in Eq.~\eqref{eq:FHcollint1}, the gluon collinear contribution in Eq.~\eqref{eq:Fgcollrap}, and the soft contribution in Eq.~\eqref{eq:Fsoftrap}, the ultraviolet poles in $\eta$ from rapidity divergences cancel. The only divergences that remain are double and single poles in $\epsilon$:
%================
\begin{eqnarray}
\label{eq:Fcollsoftreg}
\mathcal{F}^{\text{LP}}_{H\,\text{coll}} +\mathcal{F}^{\text{LP}}_{g\,\text{coll}}
+\mathcal{F}^{\text{LP}}_{\text{soft}} =
\frac{g_s^2 y_b}{16\pi^2}
\left[\frac{\mu^2}{m_b^2}\right]^\epsilon
\Bigg\{ \frac{1}{\epsilon^2} - \frac{1}{\epsilon} \left( \log\frac{\tilde P.n}{\nu_1}+ \log\frac{p_3.\bar n}{\nu_3}+\log\frac{\nu^2}{m_b^2} - 2 \right)
\nonumber \\
\hspace{3cm}
-\frac{\pi^2}{6}
+ \int_{-1}^{+1}\!\!\! d\zeta \, \zeta^2 \frac{\log\big(1- (1 - \zeta^2) r^2 - i \epsilon\big)}{1 - \zeta^2} \Bigg\}.
\end{eqnarray}
%================
Upon adding the hard contribution in Eq.~\eqref{eq:F0hardeps} to get the complete LP form factor, the double poles in $\epsilon$ cancel:
%================
\begin{eqnarray}
\label{eq:Fhardcollsoft}
\mathcal{F}^{\text{LP}} &=&
\frac{g_s^2 y_b}{16\pi^2}
\left[\frac{\mu^2}{m_b^2}\right]^\epsilon
\Bigg\{ - \frac{1}{\epsilon} \left( \log\frac{\tilde P.n\, p_3.\bar n}{-\hat s - i \epsilon}+\log\frac{\nu^2}{\nu_1\, \nu_3}  \right)
-\frac12 \log^2 \frac{-\hat s - i \epsilon}{m_b^2}  + 2 \log \frac{-\hat s - i \epsilon}{m_b^2}
\nonumber \\
&&\hspace{3cm}
-6 + \int_{-1}^{+1}\!\!\! d\zeta \, \zeta^2 \frac{\log\big(1- (1 - \zeta^2) r^2 - i \epsilon\big)}{1 - \zeta^2} \Bigg\}.
\end{eqnarray}
%================
Since $\tilde P.n\, p_3.\bar n = \hat s$, the single poles in $\epsilon$  cancel provided the rapidity regularization scales satisfy
%================
\begin{equation}
\label{eq:rapidregscales}
\nu_1\, \nu_3 = e^{+i \pi}\,  \nu^2.
\end{equation}
%================
In  Ref.~\cite{Braaten:2017lxx}, this nontrivial constraint on the rapidity regularization scales was verified by comparing with the result from analytic regularization. It would be preferable to derive it from deeper theoretical considerations. The final result for the LP form factor is obtained by inserting the integral in Eq.~\eqref{eq:intzeta} into Eq.~\eqref{eq:Fhardcollsoft}.  It agrees with the result in Eq.~\eqref{eq:FFLP-b} obtained from Refs.~\cite{Ellis:1987xu,Baur:1989cm}.

%%%%%%%%%%%%%%%%%%%%%%%%%%%%%%%%%%%%%%%%%%%
\section{Renormalized Factorization Formula}
\label{sec:RenormalizedFF}
%%%%%%%%%%%%%%%%%%%%%%%%%%%%%%%%%%%%%%%%%%%

In this Section, we present a renormalized factorization formula for the LP form factor, in which the ultraviolet divergences in each of the regularized pieces is removed by the minimal subtraction of the poles from dimensional regularization and from rapidity regularization. We also define an LP$b$ form factor in which the errors are reduced from order $m_H^2/Q^2$ to order $m_b^2/Q^2$.

%%%%%%%%%%%%%%%%%%%%%%%%%%%%%%%%
\subsection{LP form factor}
%%%%%%%%%%%%%%%%%%%%%%%%%%%%%%%%

The factorization formula for the LP form factor is given in a schematic form in Eq.~\eqref{eq:Ffact}. The explicit form of the renormalized factorization formula is 
%================
\begin{eqnarray}
\mathcal{F}^{\text{LP}}(\hat s, m_b^2,m_H^2) &\equiv& \widetilde{\mathcal{F}}_{H+g}(\hat s)
+ \int_{-1}^{+1}\!\!\!\!d \zeta\,  \widetilde{\mathcal{F}}_{b \bar b_{1V}+g}(\zeta) \, d_{b \bar b_{1V} \to H}(\zeta;m_b^2,m_H^2,P.n) 
\nonumber\\
&&\hspace{1.5cm}
+ \int_{-1}^{+1}\!\!\!\!d \zeta\,   \widetilde{\mathcal{F}}_{H+b \bar b_{8T}}(\zeta) \, d_{b \bar b_{8T} \to g}(\zeta;m_b^2,p_3.\bar n) 
+\mathcal{F}_{\text{endpt}}(m_b^2) .~
\label{eq:FfactLP}
\end{eqnarray}
%================
All the dependences on physical scales are indicated explicitly by the  arguments  in Eq.~\eqref{eq:FfactLP}. Each of the individual pieces in the factorization formula is given below.

The regularized hard contribution to the LP form factor is given in Eq.~\eqref{eq:F0hardeps}. We define the renormalized contribution from direct production of $H+g$ by minimal subtraction of the poles in $\epsilon$:
%================
\begin{equation}
\widetilde{\mathcal{F}}_{H+g}(\hat s) = 
\frac{g_s^2 y_b}{16 \pi^2} 
\left(  -\frac12 \log^2\frac{-\hat s - i \epsilon}{\mu^2}
 +2\log\frac{-\hat s - i \epsilon}{\mu^2} + \frac{\pi^2}{6}   - 6 \right).
\label{eq:Fhardren}
\end{equation}
%================
With the measure of the dimensionally regularized momentum integral defined in Eq.~\eqref{eq:intqD}, the minimal subtraction of the poles in $\epsilon$ corresponds to the modified minimal subtraction ($\overline{\text{MS}}$) renormalization scheme. The renormalized  hard contribution  depends logarithmically on $\hat s$.

The Higgs collinear contribution to the LP form factor is given by the integral over the relative longitudinal momentum fraction $\zeta$ in Eq.~\eqref{eq:FLPHcoll1V}. The hard form factor for $g^* \to b \bar b_{1V} + g$  is given in Eq.~\eqref{eq:FFbb1V+g}. The distribution amplitude for $b \bar b_{1V} \to H$ with rapidity regularization is given in Eq.~\eqref{eq:dfragbb1V->Hreg}. We define a renormalized distribution amplitude by minimal subtraction of the ultraviolet poles  in $\eta$ and in $\epsilon$:
%================
\begin{eqnarray}
d_{b \bar b_{1V} \to H}(\zeta) =
\frac{y_b}{32 \pi^2} \zeta
\left[ \log\frac{\mu^2}{m_b^2} \left(\log\frac{P.n}{\nu_1}\,\delta(1-\zeta^2)
+ \frac{1}{(1-\zeta^2)_+} \right) \right.
\nonumber\\
 \left.
 -\frac{\log\big(1-(1-\zeta^2)r^2 - i \epsilon \big)}{1-\zeta^2}\right],
 \label{eq:dbb1V->Hren}
\end{eqnarray}
%================
where $r = m_H/(2 m_b)$. The distribution amplitude depends logarithmically on $P.n$. The first integral over $\zeta$ in Eq.~\eqref{eq:FfactLP} can be evaluated by inserting the regularized distribution amplitude in Eq.~\eqref{eq:dbb1V->Hren}. The result agrees with that obtained by the minimal subtraction of the poles in the regularized integral in Eq.~\eqref{eq:FHcollint1}.

The gluon collinear contribution to the LP form factor is given by the  integral over $\zeta$ in Eq.~\eqref{eq:FLPgcoll8T}. The form factor for $g^* \to H +  b \bar b_{8T}$  is given in Eq.~\eqref{eq:FFH+bb8T}. The distribution amplitude  for $b \bar b_{8T} \to g$ with rapidity regularization is given in Eq.~\eqref{eq:dfragtt8T->greg}. We define a renormalized distribution amplitude  by minimal subtraction of the ultraviolet poles in $\eta$ and in $\epsilon$:
%================
\begin{equation}
d_{b \bar b_{8T} \to g}(\zeta) = 
\frac{g_s}{32 \pi^2}
\log\frac{\mu^2}{m_b^2}
\left( \log\frac{p_3.\bar n}{\nu_3} \delta(1-\zeta^2)+\frac{1}{(1-\zeta^2)_+}\right).
 \label{eq:dbb8T->gren}
\end{equation}
%================
The distribution amplitude depends logarithmically on $p_3.\bar n$. The second integral over $\zeta$ in Eq.~\eqref{eq:FfactLP} can be evaluated by inserting the regularized distribution amplitude in Eq.~\eqref{eq:dbb8T->gren}. The result agrees with that obtained by the minimal subtraction of the poles in the regularized integral in Eq.~\eqref{eq:Fgcollrap}.

The soft contribution to the LP form factor using rapidity regularization is given in Eq.~\eqref{eq:Fsoftrap}. We define the renormalized endpoint contribution by minimal subtraction of the ultraviolet poles in $\eta$ and in $\epsilon$:
%================
\begin{equation}
\label{eq:Fsoftren}
\mathcal{F}_{\text{endpt}}(m_b^2) =
\frac{g_s^2 y_b}{16\pi^2}
\left( \frac12 \log^2\frac{\mu^2}{m_b^2} - \log\frac{\mu^2}{m_b^2}\log\frac{\nu^2}{m_b^2} - \frac{\pi^2}{6} \right).
\end{equation}
%================
The endpoint contribution depends logarithmically on $m_b$.

The sum of the four terms in Eq.~\eqref{eq:FfactLP} reproduces the LP form factor in Eq.~\eqref{eq:FFLP-b}. The logarithms of $P.n$ from the Higgs collinear term and $p_3.\bar n$ from the gluon collinear term combine to give a logarithm of $\hat s$. The last three terms in Eq.~\eqref{eq:FfactLP} depend on the rapidity regularization scales $\nu_1$, $\nu_3$, and $\nu$. The dependence on these scales cancels upon using the relation between $\nu_1$, $\nu_3$, and $\nu$ in Eq.~\eqref{eq:rapidregscales}. All four terms in Eq.~\eqref{eq:FfactLP} depend on the dimensional regularization scale $\mu$. The dependence on $\mu$ cancels when the four terms are added.

%%%%%%%%%%%%%%%%%%%%%%%%%%%%%%%%
\subsection{Improved mass dependence}
%%%%%%%%%%%%%%%%%%%%%%%%%%%%%%%%

The leading errors in the LP form factor come from the omission of terms in the form factor that are suppressed either by $m_b^2/Q^2$ or by $m_H^2/Q^2$. Since $m_H$ is an order of magnitude larger than $2m_b$, one should be able to improve the accuracy by keeping the leading terms of the expansion in $m_b^2/Q^2$ without expanding  in $m_H^2/Q^2$. This will not change the parametric dependence of the error, which still decreases as $1/Q^2$, but one might hope for a decrease in the numerical size of the error by two orders of magnitude. We refer to the leading term in the expansion of the form factor in powers of $m_b^2/Q^2$ as the {\it LP$b$ form factor}, and we denote it by $\mathcal{F}^{\text{LP}b}(\hat s, m_b^2,m_H^2)$. It can be defined as the leading power in $M^2/Q^2$, where the hard scale is $Q\sim P_T, \sqrt{\hat{s}}$ and the soft scale is $M\sim m_b$, but $m_H$ is an arbitrary intermediate scale. The LP$b$ form factor includes terms of all powers in $m_H^2/Q^2$ that are not suppressed by $m_b^2/Q^2$, and therefore has an error of order $m_b^2/Q^2$. We will show that it can be expressed in the same form as the factorization formula for the LP form factor in Eq.~\eqref{eq:FfactLP}, with the only change being that the hard form factor $\widetilde{\mathcal{F}}_{H+g}(\hat s)$ is replaced by an $m_H$-dependent hard form factor $\widetilde{\mathcal{F}}_{H+g}^{(H)}(\hat s, m_H^2)$. The schematic form of the LP$b$  factorization formula  is 
%================
\begin{eqnarray}
\mathcal{F}^{\text{LP}b}[H+g] &\equiv&  \widetilde{\mathcal{F}}^{(H)} [H+g]   
+ \widetilde{\mathcal{F}}[b \bar b_{1V}+g]  \otimes d[b \bar b_{1V} \to H]
\nonumber\\
&& \hspace{2cm}
+ \widetilde{\mathcal{F}} [H + b \bar b_{8T}]  \otimes d[b \bar b_{8T} \to g]
+ \mathcal{F}_\text{endpt}[H+g].
\label{eq:FfactLPb}
\end{eqnarray}
%================

We first explain how the $m_H$-dependent hard form factor $\widetilde{\mathcal{F}}^{(H)} [H+g]$ in Eq.~\eqref{eq:FfactLPb} can be obtained. The schematic factorization formula for the LP form factor $\mathcal{F}^\text{LP}[H+g]$ in Eq.~\eqref{eq:Ffact} can be solved for the hard form factor $ \widetilde{\mathcal{F}}[H+g]$. Since this hard form factor does not depend on $m_H$ or $m_b$, it can be expressed as a double limit as $m_b\to 0$ and $m_H\to 0$:
%=================
\begin{eqnarray}
 \widetilde{\mathcal{F}}[H+g]  &=& \Big[ \mathcal{F}^\text{LP}[H+g]  
- \widetilde{\mathcal{F}}[b \bar b_{1V}+g]  \otimes d[b \bar b_{1V} \to H]
\nonumber\\
&& \hspace{2.2cm}
- \widetilde{\mathcal{F}} [H + b \bar b_{8T}]  \otimes d[b \bar b_{8T} \to g]
- \mathcal{F}_\text{endpt}[H+g] \Big]_{\substack{m_b\to 0 \\ m_H\to 0}}.
\label{eq:hardcontribution2}
\end{eqnarray}
%=================
The LP form factor differs from the full form factor only by terms with higher powers of $m_H$ and $m_b$, so $\mathcal{F}^\text{LP}[H+g]$ can be replaced by  $\mathcal{F}[H+g]$ inside the limits. We define a regularized $m_H$-dependent hard form factor by making this replacement and then removing the limit $m_H \to 0$:
%=================
\begin{eqnarray}
 \widetilde{\mathcal{F}}^{(H)}[H+g]  &\equiv&  \Big[ \mathcal{F}[H+g]
 - \widetilde{\mathcal{F}}[b \bar b_{1V}+g]  \otimes d[b \bar b_{1V} \to H]
\nonumber\\
&& \hspace{2cm}
- \widetilde{\mathcal{F}} [H + b \bar b_{8T}]  \otimes d[b \bar b_{8T} \to g]
- \mathcal{F}_\text{endpt}[H+g] \Big]_{m_b\to 0}.
\label{eq:hardcontribution4}
\end{eqnarray}
%=================
All four terms on the right side have additional infrared divergences in the limit $m_b\to 0$. The additional infrared divergences cancel in the sum of the four terms leaving the same infrared poles in $\epsilon$ as in the regularized hard form factor in Eq.~\eqref{eq:F0hardeps}.

We now proceed to show that the errors in the LP$b$ form factor defined by Eq.~\eqref{eq:FfactLPb} are order $m_b^2/Q^2$. The difference between the LP$b$ form factor and the LP form factor in Eq.~\eqref{eq:Ffact} is equal to the difference $\widetilde{\mathcal{F}}^{(H)}[H+g] -\widetilde{\mathcal{F}}[H+g]$ between the hard form factors, which is order $m_H^2/\hat s$. Since the error in the LP form factor decreases as $1/\hat s$, the error in the LP$b$ form factor must also decrease as $1/\hat s$. By inserting the expression for $\widetilde{\mathcal{F}}^{(H)}[H+g]$ in Eq.~\eqref{eq:hardcontribution4} into the expression for $\mathcal{F}^\text{LP$b$}[H+g]$ in Eq.~\eqref{eq:FfactLPb}, we find that the difference between the LP$b$ form factor and the full form factor can be expressed as
%================
\begin{eqnarray}
\mathcal{F}^\text{LP$b$}[H+g]  - \mathcal{F}[H+g] &=&  
\left(\mathcal{F}[H+g]\big|_{m_b= 0}  - \mathcal{F}[H+g] \right)
\nonumber\\
&& \hspace{0cm}
+ \widetilde{\mathcal{F}}[b \bar b_{1V}+g]  \otimes \left(d[b \bar b_{1V} \to H] - d[b \bar b_{1V} \to H]\big|_{m_b= 0} \right)
\nonumber\\
&& \hspace{0cm}
+ \widetilde{\mathcal{F}}[H+b \bar b_{8T}]  \otimes \left(d[b \bar b_{8T} \to g] - d[b \bar b_{8T} \to g]\big|_{m_b= 0} \right)
\nonumber\\
&& \hspace{0cm}
+ \Big(  \mathcal{F}_\text{endpt}[H+g] -  \mathcal{F}_\text{endpt}[H+g]\big|_{m_b= 0} \Big).
\label{eq:FFdiff}
\end{eqnarray}
%=================
Each term on the right side is 0 for $m_b=0$, so the right side is proportional to $m_b^2$. Since the error in the LP$b$ form factor decreases as $1/\hat s$, it must be order $m_b^2/\hat s$.

The additional infrared divergences in the terms on the right side of Eq.~\eqref{eq:hardcontribution4} could be regularized by the $b$ quark mass. This requires calculating all four terms on the right side of Eq.~\eqref{eq:hardcontribution4} with nonzero $m_b$ and then taking the limit $m_b \to 0$ at the end. The additional infrared divergences appear as logarithms of $m_b$, and they cancel between the four terms on the  right side of Eq.~\eqref{eq:hardcontribution4}. Calculating the full form factor $\mathcal{F}$ and then taking the limit $m_b \to 0$ still requires a calculation involving all three scales $\hat s$, $m_H$, and $m_b$. If this were necessary, the LP$b$ form factor would have no calculational advantage over the full form factor. However instead of taking the limit $m_b \to 0$, the $m_H$-dependent hard form factor can be calculated more easily by setting $m_b=0$ from the beginning in all four terms. The additional infrared divergences from setting $m_b= 0$ can be regularized using dimensional regularization. The additional poles in $\epsilon$ cancel in the sum of the four terms.

We proceed to calculate each of the four terms on the right side of Eq.~\eqref {eq:hardcontribution4} with $m_b=0$. The first term on the right side of Eq.~\eqref{eq:hardcontribution4} can be obtained by setting $m_b=0$ in the dimensionally regularized expression for the full form factor in Eq.~\eqref{eq:Fintdimreg}:
%================
\begin{eqnarray}
\mathcal{F}(\hat s, 0, m_H^2) &=& \frac{2 ig_s^2 y_b}{D-2} \int_q 
\frac{1}{[(q \!+\! P)^2   \!+\! i \epsilon] \, [q^2  \!+\! i \epsilon]\, [(q \!-\! p_3)^2   \!+\! i \epsilon]}.
\nonumber\\
&&\hspace{-2cm}\times \left(
(5-D)q^2 - 4\frac{(P+p_3).q\, p_3.q}{P.p_3}
+  2(D-3)p_3.q + (D-2)P.p_3 
\right).
\label{eq:Fintfull}
\end{eqnarray}
%================
After evaluating the loop integral, we get
%================
\begin{eqnarray}
\mathcal{F}(\hat s, 0, m_H^2) =
\frac{g_s^2 y_b}{16 \pi^2} \left[\frac{\mu^2}{m_H^2}\right]^\epsilon   e^{i \pi \epsilon} 
\left[-\frac{1}{2}\log^2\frac{\hat s+ i \epsilon}{m_H^2}
+ \left(\frac{1}{\epsilon} +\frac{2\hat{s}}{\hat{s}-m_H^2} \right) \log\frac{\hat s+ i \epsilon}{m_H^2}  
 -2  \right].
\label{eq:Ffull.mb0}
\end{eqnarray}
%================
The pole in $\epsilon$ has an infrared origin.

In the second term on the right side of Eq.~\eqref{eq:hardcontribution4}, all the dependence on $m_b$  is in the function $d(\zeta)$ in the expression for the distribution amplitude $d_{b \bar b_{1V} \to H}(\zeta)$  in Eq.~\eqref{eq:dfragbb1V->H}. The expression for $d(\zeta)$ is obtained by setting $m_b=0$ in Eq.~\eqref{eq:drapid}:
%================
\begin{equation}
d(\zeta) = -i  \int_q  
\frac{ \delta(\zeta - 2q.n/P.n) }
{[(\tfrac12 P+q)^2  + i \epsilon]\, [(\tfrac12 P-q)^2   + i \epsilon] }.
\label{eq:drapidmb=0}
\end{equation}
%================
The integral has no rapidity divergences, so dimensional regularization is sufficient. The analytic expression for the integral is
%================
\begin{equation}
d(\zeta) = \frac{1}{32\pi^2 \epsilon}
\left[\frac{\mu^2}{m_H^2} \right]^\epsilon
\left( e^{-i \pi} \frac{1 - \zeta^2}{4} \right)^{-\epsilon}.
 \label{eq:dmb=0}
\end{equation}
%================
The poles in $\epsilon$ can be  made explicit by using the expansion
%================
\begin{eqnarray}
\frac{1}{1 - \zeta^2} \bigg( \frac{1 - \zeta^2}{4} \bigg)^{-\epsilon} &=&
- \frac{\Gamma^2(1-\epsilon)}{\epsilon
\, \Gamma(1-2\epsilon)} \, \delta\big(1 - \zeta^2\big) + \frac{1}{(1-\zeta^2)_+} 
\nonumber\\
&& \hspace{1cm}
- \epsilon \bigg( \frac{\log(1-\zeta^2)-2\log 2}{1-\zeta^2}\bigg)_{\!\!\!+} + O(\epsilon^2).
 \label{eq:expandplus}
\end{eqnarray}
%================
The resulting expression for the regularized distribution amplitude in Eq.~\eqref{eq:dfragbb1V->H} is
%================
\begin{eqnarray}
d_{b \bar b_{1V} \to H}(\zeta)\Big|_{m_b=0} &=&
\frac{y_b\,\zeta}{32\pi^2}
\left[\frac{\mu^2}{m_H^2}\right]^\epsilon e^{i \pi \epsilon}
\left\{\left(-\frac{1}{\epsilon^2}+\frac{\pi^2}{6}\right)\delta(1-\zeta^2)
+\frac{1}{\epsilon}\frac{1}{(1-\zeta^2)_+}\right.
\nonumber\\
&&\hspace{4cm}
\left.-\left(\frac{\log(1-\zeta^2)-2\log 2}{1-\zeta^2}\right)_+
\right\}.
\end{eqnarray}
%================
Multiplying by  the hard form factor in Eq.~\eqref{eq:FFbb1V+g} and integrating over $\zeta$, we obtain
%================
\begin{equation}
\label{eq:Hcol.mb0}
\int_{-1}^{+1}\!\!\!\!d \zeta\,  \widetilde{\mathcal{F}}_{b \bar b_{1V}+g}(\zeta) \, d_{b \bar b_{1V} \to H}(\zeta)\Big|_{m_b=0} =
\frac{g_s^2y_b}{16\pi^2} \left[\frac{\mu^2}{m_H^2}\right]^\epsilon   e^{i \pi \epsilon}
\left(\frac{1}{\epsilon^2}+\frac{2}{\epsilon}-\frac{\pi^2}{6} + 4 \right).
\end{equation}
%================

In the third term on the right side of Eq.~\eqref{eq:hardcontribution4}, all the dependence on $m_b$ is in the function $d_0(\zeta)$ in the expression for the distribution amplitude $d_{b \bar b_{8T} \to g}(\zeta)$ in Eq.~\eqref{eq:dfragtt8T->g}. The expression for $d_0(\zeta)$ is obtained by setting $m_b=0$ in Eq.~\eqref{eq:d0rapid}. The resulting dimensionally regularized integral is 0, because it has no scale. The last term on the right side of Eq.~\eqref{eq:hardcontribution4} is given by the integral in Eq.~\eqref{eq:Fsoftrapint} with $m_b=0$. The three denominators are proportional to $q.\bar n$, $q^2$ and $q.n$. This dimensionally regularized integral is also zero, because it has no scale. Thus the only nonzero terms on the right side of Eq.~\eqref{eq:hardcontribution4} with $m_b=0$ are the first and second terms, which are given in Eq.~\eqref{eq:Ffull.mb0} and \eqref{eq:Hcol.mb0}.

Subtracting Eq.~\eqref{eq:Hcol.mb0} from Eq.~\eqref{eq:Ffull.mb0}, we get the regularized $m_H$-dependent hard form factor:
%================
\begin{equation}
\mathcal{F}^{\text{LP}b}_\text{hard}(\hat s, m_H^2)= 
- \frac{g_s^2 y_b}{16 \pi^2} \left[\frac{-\hat s - i \epsilon}{\mu^2}\right]^{-\epsilon}
\left(  \frac{1}{\epsilon^2} +\frac{2}{\epsilon}  - \frac{\pi^2}{6}  + 6
-\frac{2m_H^2}{\hat{s}-m_H^2}\log\frac{\hat{s}+ i \epsilon}{m_H^2} \right).
\label{eq:F0hardepsmH}
\end{equation}
%================
This reduces to the hard form factor in Eq.~\eqref{eq:F0hardeps} in the limit $m_H\to 0$. The renormalized $m_H$-dependent hard form factor can be obtained  by minimal subtraction of the poles in $\epsilon$:
%================
\begin{equation}
\widetilde{\mathcal{F}}_{H+g}^{(H)}(\hat s, m_H^2) = 
\frac{g_s^2 y_b}{16 \pi^2} 
\left( - \frac12 \log^2\frac{-\hat s - i \epsilon}{\mu^2}
 + 2\log\frac{-\hat s - i \epsilon}{\mu^2} +  \frac{\pi^2}{6} - 6
+ \frac{2m_H^2}{\hat{s}-m_H^2}\log \frac{\hat{s}+i \epsilon}{m_H^2}    \right).
\label{eq:FhardrenmH}
\end{equation}
%================

The explicit form of the LP$b$  factorization formula   in Eq.~\eqref{eq:FfactLPb}, which includes all terms at leading power of $m_b^2/Q^2$, is
%================
\begin{eqnarray}
\mathcal{F}^{\text{LP}b}(\hat s, m_b^2,m_H^2) &\equiv& \widetilde{\mathcal{F}}_{H+g}^{(H)}(\hat s,m_H^2)
+ \int_{-1}^{+1}\!\!\!\!d \zeta\,  \widetilde{\mathcal{F}}_{b \bar b_{1V}+g}(\zeta) \, d_{b \bar b_{1V} \to H}(\zeta;m_b^2,m_H^2,P.n) 
\nonumber\\
&&\hspace{1.5cm}
+ \int_{-1}^{+1}\!\!\!\!d \zeta\,   \widetilde{\mathcal{F}}_{H+b \bar b_{8T}}(\zeta) \, d_{b \bar b_{8T} \to g}(\zeta;m_b^2,p_3.\bar n) 
+\mathcal{F}_{\text{endpt}}(m_b^2) .~
\label{eq:FFLPb}
\end{eqnarray}
%================
It differs from the LP form factor in Eq.~\eqref{eq:FFLP-b} only by the difference between the $m_H$-dependent hard form factor in Eq.~\eqref{eq:FhardrenmH} and the hard form factor in Eq.~\eqref{eq:Fhardren}:
%================
\begin{equation}
\mathcal{F}^{\text{LP}b}(\hat s, m_b^2, m_H^2) = \mathcal{F}^\text{LP}(\hat s, m_b^2, m_H^2)
+\frac{g_s^2 y_b}{32\pi^2} \left\{ \frac{4m_H^2}{\hat{s}-m_H^2}\log\frac{\hat{s}+i \epsilon}{m_H^2} \right\}.
\label{eq:FFLPb-mH}
\end{equation}
%================

%%%%%%%%%%%%%%%%%%%%%%%%%%%%%%%%%%%%%%%%%%%
\section{Comparison with full form factor}
\label{sec:compare}
%%%%%%%%%%%%%%%%%%%%%%%%%%%%%%%%%%%%%%%%%%%

We proceed to compare our  approximations to the form factor at LO for $q\bar{q}\to H+g$ from the bottom-quark loop. The full form factor $\mathcal{F}(\hat s,m_b^2,m_H^2)$ is given in Refs.~\cite{Ellis:1987xu, Baur:1989cm}.
The three approximations are 
%%%%%%%%%%%%%
\begin{itemize}
\item
the {\it LP form factor} $\mathcal{F}^\text{LP}(\hat s,m_b^2,m_H^2)$ in Eq.~\eqref{eq:FFLP-b}, which is leading power in $m_H^2/\hat s$ and $m_b^2/\hat s$,
\item
the {\it LP$b$ form factor}  $\mathcal{F}^{\text{LP}b}(\hat s,m_b^2,m_H^2)$  in Eq.~\eqref{eq:FFLPb-mH}, which is leading power in $m_b^2/\hat s$ only,
\item
the $m_b \to 0$ form factor $\mathcal{F}(\hat s,m_b^2\to 0,m_H^2)$ in Eq.~\eqref{eq:FFmb->0}, which is obtained from the full form factor by taking the limit $m_b \to 0$ except in logarithms of $m_b$.
\end{itemize}
%%%%%%%%%%%%%
The full form factor and the approximations depend on the coupling constants $g_s$ and $y_b= m_b/v$ and on the masses $m_H$ and $m_b$. The mass of the Higgs is $m_H=125$~GeV. For the bottom-quark mass $m_b$, we use the running mass at the scale of the Higgs mass: $m_b(m_H)=3.06$ GeV.

The form factor ${\cal F}$ at LO has an overall coupling-constant factor $g_s^2 \,y_b$. The coupling constants depend logarithmically on momentum scales that can range from $m_b$ to $m_H$ and to $\sqrt{\hat s}$. One advantage of the LP factorization formula is that the separation of scales allows the momentum scales of the coupling constants to be determined. In the hard form factor in Eq.~\eqref{eq:Fhardren}, the appropriate scale is $\sqrt{\hat s}$, so the coupling constant factor is $g_s^2(\sqrt{\hat s}\,) \,y_b(\sqrt{\hat s}\,)$. In the Higgs collinear contribution, the coupling constant factor in the hard form factor for $g^* \to b \bar b_{1V} + g$ in Eq.~\eqref{eq:FFbb1V+g} is $g_s^2(\sqrt{\hat s}\,)$, and the Yukawa coupling constant in the distribution amplitude for $ b \bar b_{1V} \to H$ in Eq.~\eqref{eq:dbb1V->Hren} is $y_b(m_H)$. In the gluon collinear contribution,  the coupling constant factor in the hard form factor for  $g^* \to H +  b \bar b_{8T}$ in Eq.~\eqref{eq:FFH+bb8T}  is $g_s(\sqrt{\hat s}\,) \,y_b(\sqrt{\hat s}\,)$, and the coupling constant in the distribution amplitude for $b \bar b_{8T} \to g$ in Eq.~\eqref{eq:dbb8T->gren} is $g_s(m_b)$. In the endpoint contribution in Eq.~\eqref{eq:Fsoftren}, the coupling constant in the hard amplitude for producing $b + \bar b$ is $g_s(\sqrt{\hat s}\,)$, and the coupling-constant factor in the soft amplitude for the transition $ b \bar b \to H+g$ is  $g_s(m_b) \,y_b(m_H)$. Replacing the coupling constants by the appropriate running coupling constants resums some leading logarithms to all orders. However there is little to be gained by this partial resummation unless all the leading logarithms are summed to all orders. If we choose the common scale $m_H$ for all the coupling constants, the coupling constant factor $g_s^4 y_b^2$ in $|{\cal F}|^2$ is approximately $2.4 \times 10^{-4}$. We will compare the approximations to $|{\cal F}|^2$ in ways that do not depend on the coupling constant factor.

%%%%%%%%%%%%%
\begin{figure}
\includegraphics[width=12cm]{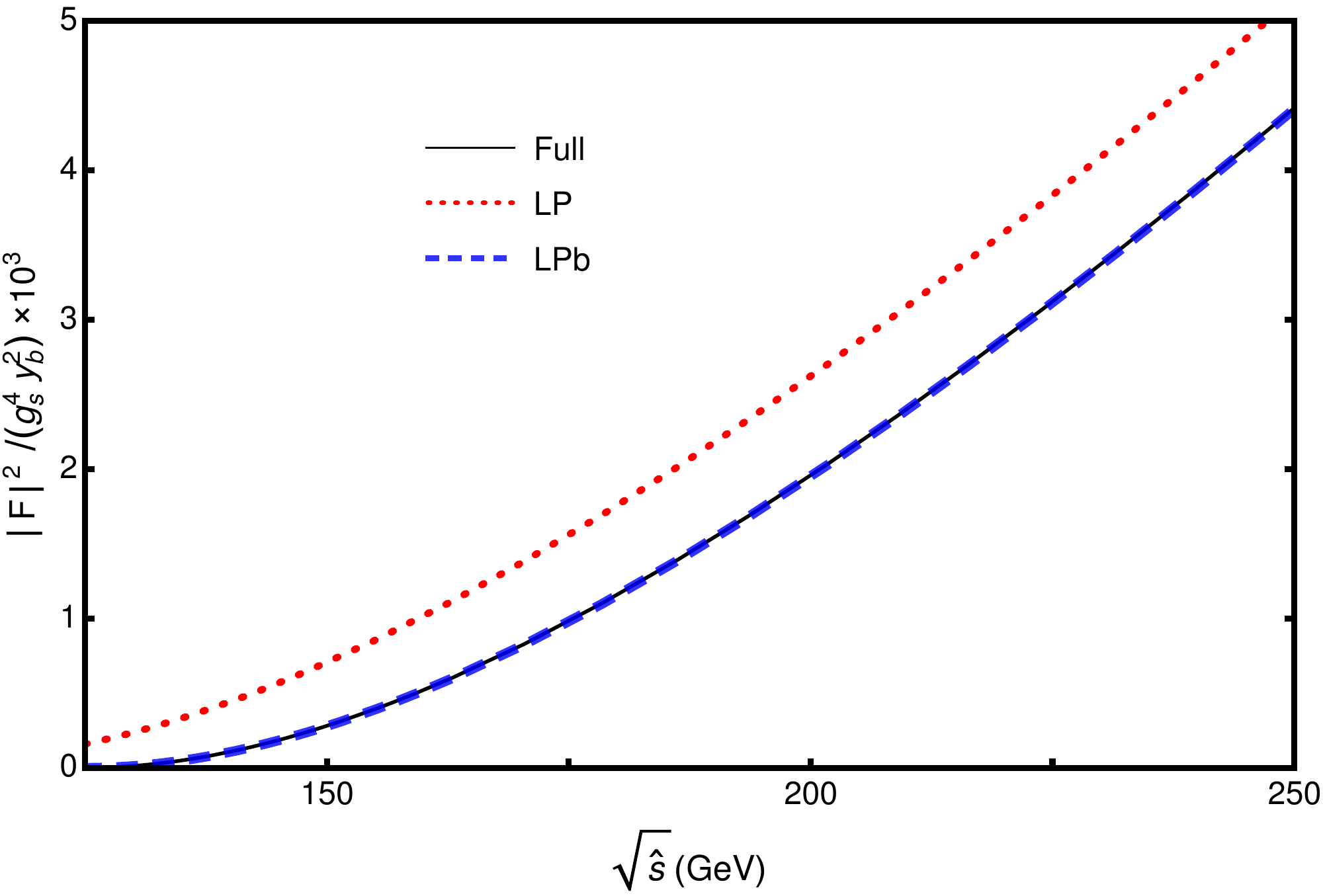}
\caption{Form factors for $q\bar{q}\to H+g$ with a bottom-quark loop as functions of the center-of-mass energy $\sqrt{\hat{s}}$: the full form factor $|\mathcal{F}|^2$ (solid curve), the LP form factor  (dotted curve), and the LP$b$ form factor (dashed curve). The $m_b \to 0$ form factor is indistinguishable from the full form factor in this plot.
\label{fig:bFullvsLP}}
\centering
\end{figure}
%%%%%%%%%%%%%

In Fig.~\ref{fig:bFullvsLP}, the absolute squares of the form factors divided by the coupling constant factor $g_s^4 y_b^2$ are shown as functions of the center-of-mass energy $\sqrt{\hat{s}}$, which ranges from the threshold $m_H$ for producing the Higgs to 250~GeV. The $m_b \to 0$ form factor is indistinguishable from the full form factor in this plot. The LP$b$ form factor is almost indistinguishable from the full form factor. The error in the LP form factor does not seem to be decreasing as $\hat s$ increases.

%%%%%%%%%%%%%
\begin{figure}
\includegraphics[width=7.74cm]{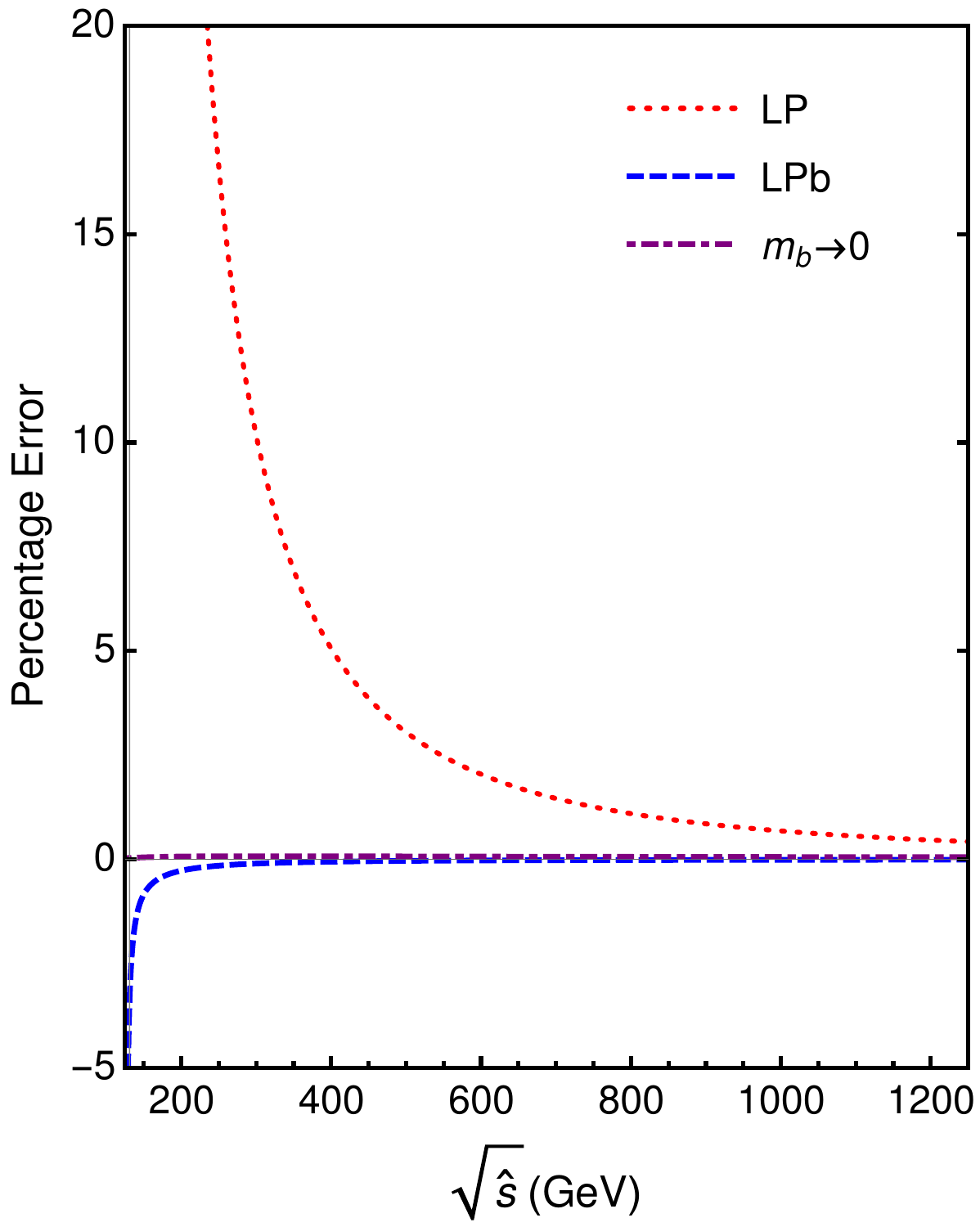} ~
\includegraphics[width=8.0cm]{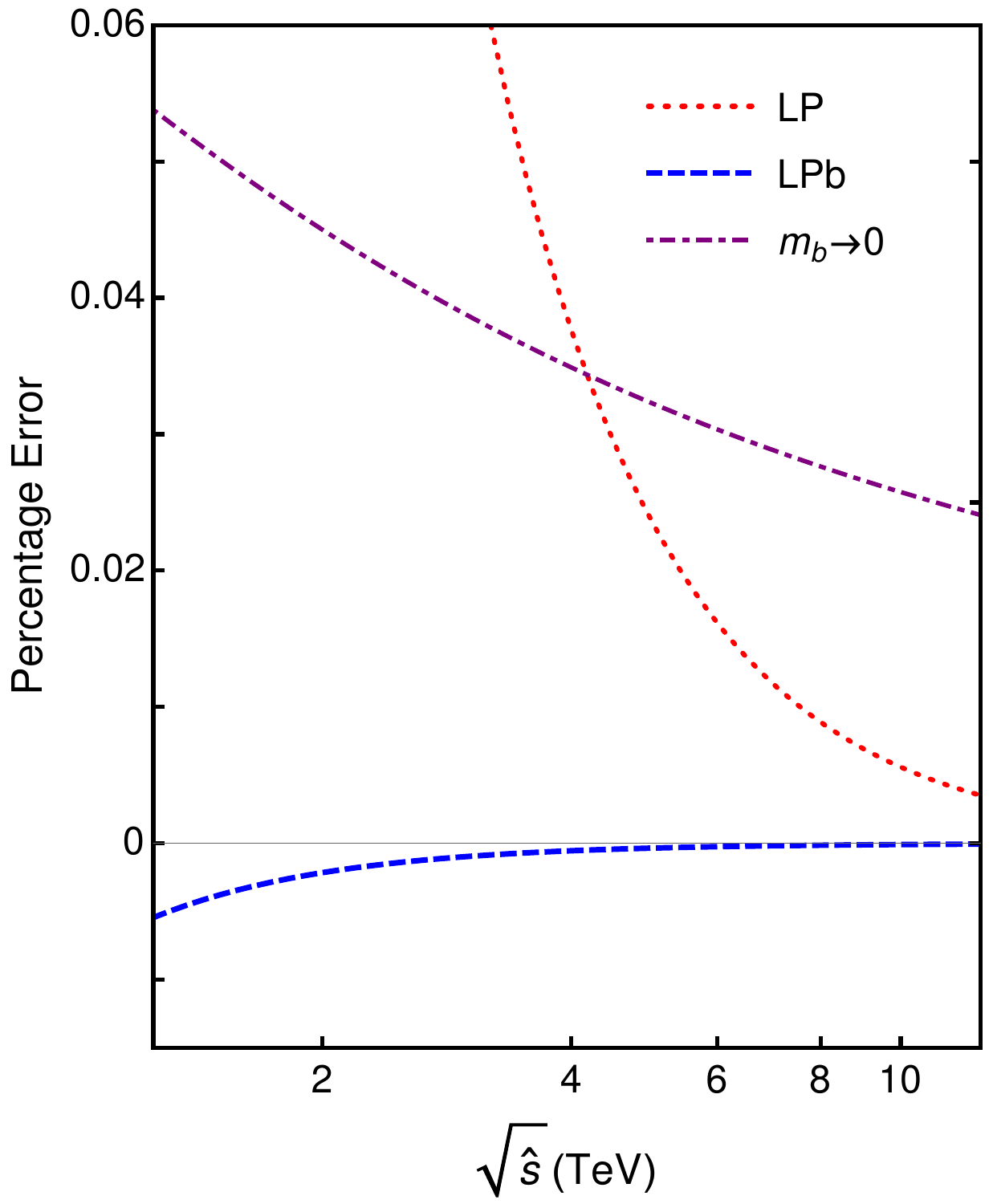}
\caption{Percentage errors in form factors for $q\bar{q}\to H+g$ with a bottom-quark loop as functions of the center-of-mass energy $\sqrt{\hat{s}}$: the LP form factor (dotted curve), the LP$b$ form factor (dashed curve), and the $m_b \to 0$ form factor (dash-dotted curve). The ranges of $\sqrt{\hat s}$ are from $m_H$ to $10\,m_H$ on a linear scale (left panel) and from $10\,m_H$ to $100\,m_H$ on a log scale (right panel). 
\label{fig:berr}}
\centering
\end{figure}
%%%%%%%%%%%%%

In Fig.~\ref{fig:berr}, we compare the percentage errors in the  three approximations to $|\mathcal{F}|^2$. The percentage error is defined as the difference from $|\mathcal{F}|^2$ divided by $|\mathcal{F}|^2$. In the left panel of Fig.~\ref{fig:berr}, which is linear in $\sqrt{\hat s}$ out to $10\, m_H$, all three approximations appear to be converging to $|\mathcal{F}|^2$ as $\hat{s}$ increases but with very different rates of convergence. The percentage error in the LP form factor approaches 0 from below, decreasing to less than 5\% for  $\sqrt{\hat{s}}$ greater than about 400~GeV. The LP$b$ form factor is a much better approximation. The percentage error approaches 0 from below, and it is already less than 5\% at $\sqrt{\hat{s}}=129$~GeV. The $m_b \to 0$ form factor seems to be a much better approximation than the LP$b$ form factor. The percentage error seems to be approaching 0 from above, and it is always less than 0.08\%. In the right panel of Fig.~\ref{fig:berr}, which is logarithmic in $\sqrt{\hat s}$ from $10\, m_H$ to $100\, m_H$, all three approximations seem to continue converging to $|\mathcal{F}|^2$. However the rates of convergence of the LP$b$ form factor and the LP form factor are much more rapid, consistent with errors that scale as $1/\hat s$. The LP$b$ form factor is a better approximation than the $m_b \to 0$ form factor for $\sqrt{\hat s}$ greater than about 330~GeV. The LP form factor is a better approximation than the $m_b \to 0$ form factor for $\sqrt{\hat s}$ greater than about 4.1~TeV. The error in the $m_b \to 0$ form factor actually approaches a nonzero constant at large $\hat s$. The leading term in the expansion of the error in powers of $m_b^2/m_H^2$ is
%================
\begin{equation}
\mathcal{F}(\hat s, m_b^2\to 0, m_H^2) - \mathcal{F}(\hat s, m_b^2, m_H^2)
\longrightarrow \frac{g_s^2 y_b m_b^2}{8\pi^2m_H^2} \left(-\log \frac{m_H^2}{m_b^2} -2 + i \pi \right).
%\label{eq:FFb0-Full}
\end{equation}
%================
The error is small because it is suppressed by a factor of $m_b^2/m_H^2$. In the right panel of Fig.~\ref{fig:berr}, the decrease in the percentage error for the $m_b \to 0$ form factor as $\hat s$ increases is due to the denominator $|\mathcal{F}|^2$ increasing as $\log^4(\hat s/m_H^2)$.

%%%%%%%%%%%%%%%%%%%%%%%%%%%%%%%%%%%%%%%%%%%
\section{Summary and Outlook}
\label{sec:Discuss}
%%%%%%%%%%%%%%%%%%%%%%%%%%%%%%%%%%%%%%%%%%%

In this work, we applied factorization methods developed for QCD to the amplitude for Higgs production at large transverse momentum through a bottom-quark loop. That amplitude is complicated by multiple energy scales: the hard kinematic scales $Q \sim P_T,\hat s^{1/2}$ and the soft mass scales $M \sim m_b,m_H$. Factorization can be used to separate the scales $M$ and $Q$ and expand the amplitude in powers of $M^2/Q^2$. To illustrate the factorization approach, we applied it to the bottom-quark-loop contribution to the amplitude for the parton process $q \bar q \to H + g$ at LO in $\alpha_s$. The matrix element for this parton process is determined by the form factor $\mathcal{F}(\hat s, m_b^2,m_H^2)$ defined in Eq.~\eqref{eq:Fdef}. We defined the {\it leading-power} (LP) form factor $\mathcal{F}^\text{LP}$ to be the leading terms in the expansion of $\mathcal{F}$ in powers of $M^2/Q^2$. A factorization formula for the LP form factor in which the scales $Q$ and $M$ are separated   is given schematically in Eq.~\eqref{eq:Ffact}. The explicit renormalized form of the factorization formula is given in Eq.~\eqref{eq:FfactLP}. Each piece in the LP factorization formula was obtained through a diagrammatic calculation that involves fewer scales than in the calculation of the full form factor. We also defined the LP$b$ form factor $\mathcal{F}^\text{LP$b$}$ to be the leading terms in the expansion of $\mathcal{F}$ in powers of $m_b^2/\hat{s}$, keeping all dependence on $m_H$ that is not suppressed by $m_b^2/\hat{s}$. The LP$b$ form factor is defined by the factorization formula  in Eq.~\eqref{eq:FFLPb}, and it is given expliicitly in Eq.~\eqref{eq:FFLPb-mH}.

In the regularized form of the  factorization formula for the LP form factor given schematically in Eq.~\eqref{eq:Ffact}, each of the four terms comes from a different region of the loop momentum of the $b$ quark. The method of regions introduces rapidity divergences in addition to the infrared and ultraviolet divergences that are regularized by dimensional regularization in $4-2 \epsilon$ dimensions. We regularized the rapidity divergences using rapidity regularization with regularization parameter $\eta$. The poles in $\eta$ cancel when the last three terms in the factorization formula in Eq.~\eqref{eq:Ffact} are added. The poles in $\epsilon$ also cancel when all four terms in the factorization formula are added, provided the rapidity regularization scales satisfy the constraint in Eq.~\eqref{eq:rapidregscales}. It would be preferable to deduce this constraint from deeper theoretical considerations. With rapidity regularization and zero-bin subtraction, the poles in $\eta$ are ultraviolet divergences. The minimal subtraction of the poles in $\eta$ and in $\epsilon$ in each of the regularized pieces of the factorization formula can therefore be interpreted as a renormalization procedure.

In the renormalized factorization formula for the LP form factor in Eq.~\eqref{eq:FfactLP}, the hard scales $Q$ and the soft scales $M$ are separated. The hard contribution depends only on the hard scale $Q$, and it is given in Eq.~\eqref{eq:Fhardren}. The Higgs collinear and gluon collinear contributions were each expressed as an integral over the relative longitudinal momentum fraction $\zeta$ of a $b$ quark of the product of a form factor that depends on the hard scale $Q$ and a distribution amplitude that depends on the soft scale $M$. The integrand for the Higgs collinear contribution is the product of the hard form factor for $t \bar t_{1V}+g$  in Eq.~\eqref{eq:FFbb1V+g} and the distribution amplitude  for $t \bar t_{1V}$ in the Higgs in Eq.~\eqref{eq:dbb1V->Hren}. The integrand for the gluon collinear contribution is the product of the hard form factor for $H+t \bar t_{8T}$ in Eq.~\eqref{eq:FFH+bb8T} and the distribution amplitude for $t \bar t_{8T}$ in a real gluon in Eq.~\eqref{eq:dbb8T->gren}. The endpoint contribution to the LP form factor depends only on the scale $M$, and it is given in Eq.~\eqref{eq:Fsoftren}.

The LP form factor $\mathcal{F}^\text{LP}$ is a good approximation to the full form factor only at extremely large $\hat{s}$. The error is of order $m_H^2/\hat{s}$, so the error decreases to 0 as $\hat{s}$ increases. As shown in Fig.~\ref{fig:berr}, the percentage error in $|\mathcal{F}^\text{LP}|^2$ does not decrease to less than 5\% until $\sqrt{\hat s} > 400$~GeV. Thus the LP form factor has no practical use at LHC energies. We defined the LP$b$ form factor $\mathcal{F}^\text{LP$b$}$ by the simple modification of the factorization formula in Eq.~\eqref{eq:FFLPb}. It differs from the LP form factor only in the $m_H$-dependent hard form factor, which is given in Eq.~\eqref{eq:FFLPb-mH}.  This hard form factor can be obtained from additional calculations with $m_b=0$. The error in the LP$b$ form factor is order $m_b^2/\hat{s}$. As shown in Fig.~\ref{fig:berr}, the percentage error in $|\mathcal{F}^\text{LP$b$}|^2$ 
is already less than $5\%$ at $\sqrt{\hat s}=129$~GeV. 

Our factorization formula for the form factor $\mathcal{F}$ for the parton process $q \bar q \to H +g$ is the sum of four terms: a hard term, two collinear terms, and an endpoint term. It can be adapted to the bottom-quark-loop contributions to the form factors for the parton processes $g\, q \to H+q$ and $g\, \bar q  \to H+\bar q$ by analytically continuing the positive Mandelstam variable $\hat s$ to a negative Mandelstam variable $\hat{t}$. The factorization formula for those form factors involves a distribution amplitude for $b \bar b$ in an incoming gluon instead of the distribution amplitude for $b \bar b$ in an outgoing gluon. The factorization formula for the bottom-quark-loop contribution for the parton process $g g \to H+g$ at LO can be derived using similar methods, but it is more complicated. It has a hard term, four collinear terms, and perhaps as many as six endpoint terms. All the pieces in the factorization formula can be obtained directly from Feynman diagrams by calculations that involve fewer scales than  the full matrix element.

Our factorization approach can be extended to NLO in $\alpha_s$. The NLO calculation of the form factor for $q \bar q \to H+g$ would require calculating each of the pieces in the factorization formula in Eq.~\eqref{eq:FfactLP} to NLO. The NLO calculations of the hard form factors $\widetilde{\mathcal{F}}_{H+g}$, $\widetilde{\mathcal{F}}_{b \bar b_{1V} +g}$, and $\widetilde{\mathcal{F}}_{H+b \bar b_{8T}}$ require straightforward perturbative QCD calculations with massless quarks and massless Higgs. The NLO calculation of the endpoint form factor $\mathcal{F}_\text{endpt}$ may also require factorization, since it may have nontrivial dependence on the scale $Q$ through the hard form factor $\widetilde{\mathcal{F}}_{b + \bar b}$ for producing $b+\bar b$. The NLO calculation of the fragmentation amplitudes  for $b \bar b_{1V} \to H$ and  for $b \bar b_{8T} \to g$ may be the most challenging steps in the NLO calculation of the LP form factor. At NLO, there may be additional terms in the factorization formula associated with other double-parton channels, such as $b \bar b_{1S}$, $b \bar b_{1T}$, $b \bar b_{8S}$, and $b \bar b_{8V}$. These additional terms would require only LO calculations.

Our factorization formula could be derived more formally using effective field theory methods analogous to those used in soft collinear effective field theory in QCD. The individual pieces in the factorization formula could all be expressed in terms of matrix elements of operators in the effective field theory. These formal definitions could be useful in the calculation of the form factor to higher orders in $\alpha_s$. They would also facilitate the all-order resummation of large logarithms of $P_T^2/m_b^2$ by solving renormalization group equations. 

We expect the LP$b$ factorization formula at NLO to be useful phenomenologically at the LHC. The effect of the bottom-quark mass on  the Higgs $P_T$ distribution at the LHC is expected to be at most $-8$\% for $P_T<50$ GeV \cite{Keung:2009bs}. The fractional error of the  LP$b$ factorization formula at LO is order $m_b^2/Q^2$. If the leading logarithms of $P_T/m_b$ can be resummed to all orders, the fractional error of the  LP$b$ factorization formula  at NLO is order $\alpha_s m_b^2/Q^2$. This error may be sufficient for phenomenological purposes at the LHC.

%======================= The End =================================
     
\acknowledgments
This work was supported in part by the Department of Energy under grant DE-SC0011726. We acknowledge the use of FeynCalc \cite{Mertig:1990an,Shtabovenko:2016sxi} in this research.

%%%%%%%%%%%%%%%%%%%%%%%%%%%%%%%%%
%%%%%%%%%%%%%%%%%%%%%%%%%%%%%%%%%
%                                                     						 
%  References				
%  											   	  
%%%%%%%%%%%%%%%%%%%%%%%%%%%%%%%%%

\providecommand{\href}[2]{#2}\begingroup\raggedright

\end{document}